\documentclass[12pt]{iopart}
\usepackage{iopams}

\usepackage{graphicx} 

\eqnobysec

\def\be{\begin{equation}} 
\def\ee{\end{equation}} 
\def\ba{\begin{eqnarray}} 
\def\ea{\end{eqnarray}} 
  
\def\ov{\overline} 
\def\I{{\rm Im}} 
\def\R{{\rm Re}} 
\def\Z{\mathbb{Z}}

\def\nl{\nonumber \\}

\def\de{\partial} 
\def\wt{\widetilde} 
\def\wh{\widehat}

\def\a{\alpha} 
\def\b{\beta} 
\def\g{\gamma} 
\def\G{\Gamma} 
\def\D{\Delta} 
\def\d{\delta} 
 
\def\eps{\varepsilon} 
\def\z{\zeta}

\def\l{\lambda} 
\def\L{\Lambda} 
\def\m{\mu}

\def\p{\pi} 
 
\def\r{\rho} 
\def\s{\sigma} 
 
\def\t{\tau} 
\def\f{\phi} 
\def\vf{\varphi} 
\def\F{\Phi} 
\def\c{\chi}

\def\winf{W_{1+\infty}} 
\def\u1{\widehat{U(1)}}
\def\su2{\widehat{SU(2)}_1}

\def\disp{\displaystyle}

\def\KK{Q}
\def\da{\downarrow}
\def\ua{\uparrow}

\begin{document} 

\title[Partition Functions of Non-Abelian Quantum Hall States]
{Partition Functions of Non-Abelian Quantum Hall States}

\author{Andrea Cappelli${}^1$ and Giovanni Viola ${}^{1,2}$}

\address{${}^1$ INFN, Via G. Sansone 1, 50019 Sesto Fiorentino (FI), Italy}
\address{${}^2$ Dipartimento di Fisica e Astronomia, 
Via G. Sansone 1, 50019 Sesto Fiorentino (FI), Italy}
\eads{\mailto{andrea.cappelli@fi.infn.it}, 
\mailto{giovanni.viola@fi.infn.it}} 

\begin{abstract}
Partition functions of edge excitations are obtained for non-Abelian Hall
states in the second Landau level, such as the anti-Read-Rezayi state, the
Bonderson-Slingerland hierarchy and the Wen non-Abelian fluid, as well as for
the non-Abelian spin-singlet state.
The derivation is straightforward and unique starting
from the non-Abelian conformal field theory data and solving the
modular invariance conditions.
The partition functions provide a complete account of the excitation 
spectrum and are used to describe experiments of Coulomb blockade
and thermopower.
\end{abstract}

\maketitle 


\section{Introduction}

Quantum Hall states \cite{wen} in the second Landau level are
intensively investigated both theoretically \cite{stern-rev}\cite{na-interf} 
and experimentally \cite{cb-exp} \cite{thermo-exp}, because they might
possess excitations with non-Abelian fractional statistics
\cite{pfaff}.  The latter could provide concrete
realizations of quantum gates that are topologically protected from
decoherence and thus implement the Topological Quantum Computation
scheme of Ref. \cite{tqc}.

Besides the original Moore-Read Pfaffian state at filling fraction
$\nu=5/2$ \cite{pfaff} and its parafermion generalization by
Read and Rezayi (RR) \cite{rr}, other non-Abelian states have been
proposed that could also explain the observed plateaux for filling
fractions $2<\nu<3$.  These are the non-Abelian spin-singlet states
(NASS), introduced in \cite{nass} and further developed in
\cite{ardonne}, the charge-conjugates of Read-Rezayi's states
($\overline{\rm RR}$) \cite{antiRR}, the Bonderson-Slingerland (BS)
hierarchy built over the Pfaffian state \cite{bs}, and Wen's $SU(n)$
non-Abelian fluids (NAF) \cite{bw}, the $SU(2)$ case in particular.

In this paper, we continue the study of partition functions for edge
excitations of quantum Hall states. Issued from conformal field theory
(CFT) data \cite{cft}, the partition functions are found in the
geometry of the annulus where they enjoy the symmetry under modular
transformations \cite{cz}: they provide a complete identification 
of the Hilbert space of excitations and can be used to describe 
experiments searching non-Abelian statistics.
Modular invariant partition functions were obtained in \cite{cgz}
\cite{cvz} for the Abelian hierarchical and non-Abelian Read-Rezayi
states; here, we provide the expressions for the other non-Abelian
states in the second Landau level.

In general, the RCFT for non-Abelian states are of the type $U(1)\times G/H$,
where the non-Abelian part is characterized by the affine symmetry group 
$G$ or the coset $G/H$ \cite{cft}, and the Abelian part $U(1)$ 
accounts for the charge of excitations.
The parameters specifying the second part (compactification radius, charges
and filling fractions) can be determined by the standard requirements
on the charge and statistics of the electron and its relative
statistics with respect to the other excitations \cite{hiera}; 
in some cases, further conditions are suggested by the
physics of the specific Hall state \cite{cgt2}.

In our analysis, the construction of partition functions of 
non-Abelian Hall states is completely straightforward. 
The inputs are: 
i) the conformal field theory $G/H$ of the neutral non-Abelian 
part of excitations, and ii) the choice of Abelian field in this theory 
representing the neutral part of the electron.
From these data, the charge and statistics of all excitations
can be self-consistently found without any further physical hypothesis.
Actually, modular invariance requires  
a non-trivial pairing between the sectors of the neutral and charged  RCFTs, 
and reproduces the standard physical conditions on the spectrum.

Modular invariance is one of the defining properties of rational
conformal field theories (RCFT); when combined with the exchange
(duality) symmetry of correlators, it implies the Moore-Seiberg
identities among $n$-point functions on Riemann surfaces \cite{mose}.
In particular, the matrix elements $S_{ab}$ of the $S$ modular
transformation of (extended) conformal characters $\theta_a$,
corresponding to the partition functions on the disk \cite{cgz},
determine the fusion rules of quasiparticles {\it via} the Verlinde
formula \cite{cft}, as well as entropy numbers describing both the
degeneracy of non-Abelian quasiparticle many-body states (quantum
dimensions $d_a=S_{a0}/S_{00}$) \cite{cft} and the bipartite
entanglement of topological fluids, ${\cal S}=\a L -\g$, where $L$ is
the length of the boundary and $\g$ is the universal term,
$\g=\frac{1}{2}\log\left(\sum_ad_a^2\right)$ \cite{tee}.

In the recent literature, the partition functions of Abelian hierarchical
and non-Abelian Read-Rezayi states were successfully employed for
describing the Coulomb Blockade (CB) current peaks \cite{cb0}, both at
zero \cite{cgz} \cite{cvz} and non-zero \cite{cbT}\cite{georgiev} temperatures.
Here we extend these analyses to the other non-Abelian
fluids; we also point out that the thermal-activated off-equilibrium
CB current can measure the degeneracies of
neutral states and distinguish between different 
Hall states with equal peak patterns at zero temperature \cite{dopp}.

Another proposed signature of non-Abelian statistics that could be
experimentally accessible \cite{thermo-exp} is the thermopower, namely the
ratio of the thermal and electric gradients at equilibrium
\cite{cooper} \cite{thermop}: this could measure the quantum dimension $d_1$ of
the basic quasiparticle in the Hall fluid.  We show that the thermopower can
be easily described by the edge partition function, with the $S$ modular
matrix playing an important role again.

Let us stress that in this paper we describe Hall states by means of 
Rational CFTs and exploit the modular invariance of their partition functions;
we do not discuss other approaches involving non-rational theories,
that could relevant  for the Jain hierarchical states in particular
\cite{cgt2} \cite{hh}.

The paper is organized as follows. In section 2, we recall
the main features of partition functions in the QHE \cite{cvz}:
we rederive their expression for the Read-Rezayi states starting from 
the non-Abelian CFT data only. In section 3, we obtain the
partition functions corresponding to the other non-Abelian states.
In section 4, we use the partition functions to compute
the Coulomb blockade current peaks at non-vanishing temperatures,
the thermopower and the associated non-Abelian entropies.
In the conclusion, we discuss further model building based on
the study of partition functions.
One Appendix contains more technical data of non-Abelian RCFT and 
modular transformations.

\section{Building partition functions}

\subsection{Modular invariance}

\begin{figure}[t]
\begin{center}
\includegraphics[width=6cm]{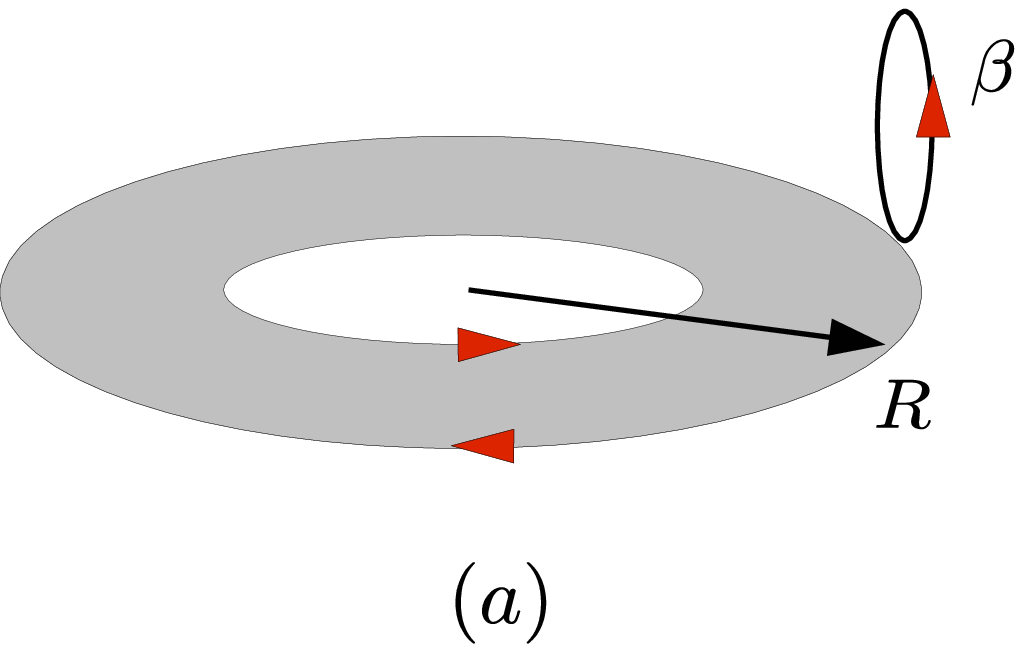}
\hspace{1cm}
\includegraphics[width=6cm]{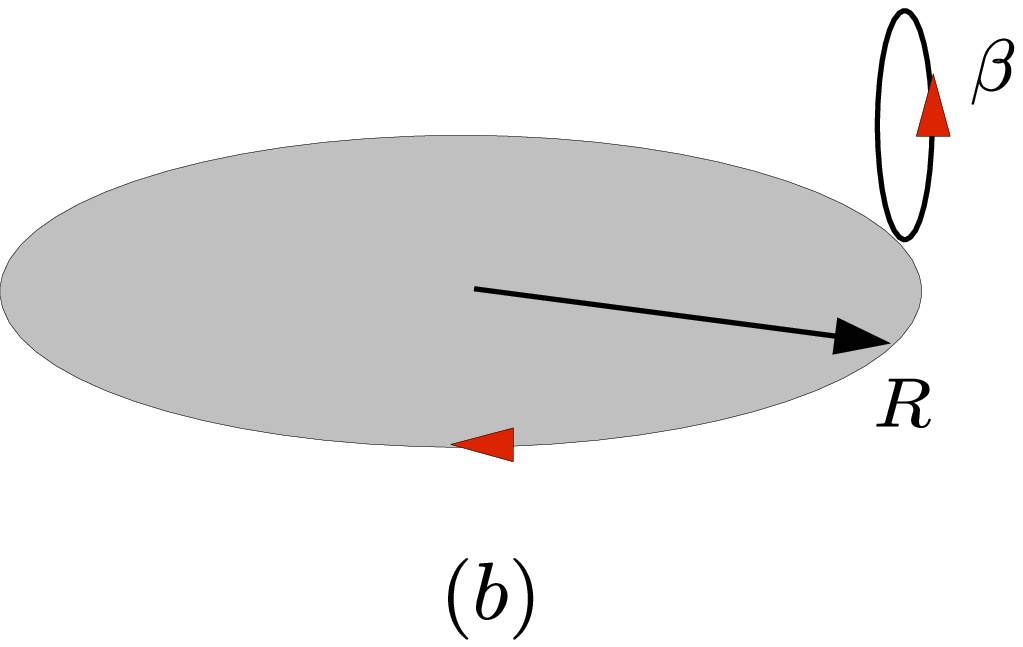}
\end{center}
\caption{The (a) annulus and (b) disk geometries.}
\label{z-geom}
\end{figure}

Partition functions are best defined for the Hall geometry of an annulus
(see Fig. \ref{z-geom}(a)), to which we add a compact Euclidean time
coordinate for the inverse temperature $\b$. This space geometry allows
for the measure of the Hall current and is equivalent to the
bar geometry (corresponding to $R\to\infty$), while enjoying 
some special symmetries. The disk geometry (Fig. \ref{z-geom}(b)), 
describing isolated Hall droplets, can be obtained from 
the annulus by shrinking the inner radius to zero.

As space-time manifold, the annulus at finite temperature has
the topology: ${\cal M}=S^1\times S^1\times I$, where $I$ is the
finite interval of the radial coordinate.  The edge
excitations live on the boundary $\partial{\cal M}$, corresponding to
two copies of a space-time torus $S^1\times S^1$: they 
are chiral and anti-chiral waves on the outer $(R)$ and inner $(L)$ edges,
respectively.
This geometry is particularly convenient owing to the symmetry under
modular transformations acting on the two periodic coordinates and
basically exchanging the space and time periods.
The partition functions should be invariant under this geometrical 
symmetry that actually implies several physical 
conditions on the spectrum of the theory \cite{cvz}.
For simplicity, we consider the case of
no bulk excitations inside the annulus: later we shall see how to include
them. 

The (grand canonical) partition function on the annulus 
is defined by \cite{cft}: 
\be
{\rm Z}_{\rm annulus}\left(\tau,\zeta \right)\ = \ {\cal K}
\ {\rm Tr}\ \left[ {\rm e}^{i2\pi \left(
\tau\left(L_0^L- c/24\right) - \ov\tau\left(L_0^R- c/24\right) +
\zeta Q^L+ \ov\zeta Q^R \right) } \right]\ ,
\label{zdef}\ee
where the trace is over the states of the Hilbert space, ${\cal K}$
is a normalization and $(\tau,\zeta)$ are complex numbers. 
The total Hamiltonian and spin are given by:
\ba
H &=& {v_R\over R_R}\left(L^R_0- {c\over 24}\right) +
{v_L\over R_L}\left(L^L_0- {c\over 24}\right)
+V_o\left(Q^L-Q^R\right) +\ {\rm const.},
\nl
J &=& L^L_0-L^R_0\ .
\label{hconf}\ea
The energy on both $L$ and $R$ edges,  $E =(v/R)(L_0 -c/24)$, is
proportional to the dilatation operator in the plane $L_0$, with eigenvalue
the conformal dimension $h$; it also includes the Casimir energy 
proportional to the Virasoro central charge $c$ \cite{cft}. 
The real and imaginary parts of $\tau$ are respectively given by
the ``torsion'' $\eta$ and the inverse temperature $\beta$ times 
the Fermi velocity $v$; the parameter $\z$ is
proportional to the chemical potential
$\mu$ and electric potential difference between the edges $V_o$:
\be
i 2 \pi\t=- \b\ \frac{v+i\eta}{R}\ ,\qquad 
i 2\pi \z=- \b\left(V_o +i\mu \right)\ .
\label{t-z-def}
\ee
Note that the charges $Q^L$ and $Q^R$ are defined in (\ref{zdef}) in such a 
way to obtain the correct coupling to the electric potential 
in (\ref{hconf}).
In the following, we shall momentarily choose a symmetric Hamiltonian 
for the two edges by adjusting the velocities of propagation of excitations,
$v_L/R_L=v_R/R_R$. 

Most theories of edge excitations in the quantum Hall effect
are rational conformal theories (RCFT) \cite{cft}:
their Hilbert space is divided into a finite number of ``sectors'', each one
describing a basic quasiparticle, with rational charge and
statistics, together with the addition of electrons; for example, 
in the Laughlin states with filling $\nu=1/p$, there are $p$ sectors,
$\l=1,\dots,p$, and the associated charge is $Q=\lambda/p + \Z$. 
In mathematical terms, each sector provides a representation of the
maximally-extended chiral symmetry algebra, which contains the Virasoro
conformal algebra as a subalgebra.

The trace over the Hilbert space in (\ref{zdef}) 
can be divided into sub-sums relative to pairs of sectors, left and right for
the inner and outer edges; the sum of states in each sector give rise to
a character $\theta_\l(\t,\z)$ of the extended algebra \cite{cft}.
As a result, the partition function reduces to 
the finite-dimensional sum:
\be
Z_{\rm annulus}\ =\ \sum_{\lambda,\ov\lambda=1}^N\
{\cal N}_{\ \l,\mu}\ \theta_\l\left(\t,\z \right) \
\ov{\theta^c_\mu\left(\t,\z \right)}\ .
\label{rcftz}\ee
In this equation, the bar denotes complex conjugation and the suffix
$(c)$ is the charge conjugation $C$, acting by: $ Q \to -Q$, $\theta \to
\theta^c$.  
The inner (resp. outer) excitations are described by
$\theta_\l$ (resp. $\theta_\mu^c$), according to the definition (\ref{zdef}). 
The coefficients ${\cal N}_{\ \l,\mu}$ are positive integer numbers giving the
multiplicities of sectors of excitations to be determined by imposing
modular invariance and some physical requirements.

In the case of the Laughlin fluids, the RCFT is an extension of the affine
$\u1$ algebra of the chiral Luttinger liquid ($c=1$).  The characters are
given by theta functions with rational characteristics \cite{cz}:
\ba
\theta_\l (\t,\z)&=&
{\rm e}^{\disp -{\pi\over p}{\left(\I\zeta\right)^2\over \I\tau} }
\ K_\l \left(\t,\z ; p\right) 
\nonumber\\
&=& {\rm e}^{\disp -{\pi\over p}{\left(\I\zeta\right)^2\over \I\tau} }
\ {1\over \eta(\t)}\ \sum_{k\in \Z}\
{\rm e}^{\disp\ i2\pi \left(\tau{(pk+\lambda)^2\over 2p} +
\zeta\left({\lambda\over p}+k\right)\right) } .
\label{thetaf}\ea
Note that each term in the summation is a character of a $\u1$ representation
with charge $Q=\lambda/p + k$, i.e. describes the addition of $k$ electrons to
the basic anyon. The Dedekind function $\eta(\t)$ in the denominator describes
particle-hole excitations and the non-analytic is as 
a measure factor; both terms are explained in \cite{cz}.

Owing to scale invariance, the doubly-periodic geometry of the torus is
specified by the modular parameter $\tau =\omega_2/\omega_1$, the ratio of the
two periods ($\I \t >0$).  The same toroidal geometry can be described by
different sets of coordinates respecting periodicity, that are related among
themselves by integer linear transformations with unit determinant.
The transformation of the modular parameter is:
\be
\tau' = \frac{a\tau +b}{c\tau +d}\ , 
\qquad a,b,c,d \in \Z\ , \qquad ad-bc = 1\ .
\label{fracl}
\ee
Modular transformations belong to the infinite discrete group 
$\Gamma \equiv PSL(2,\Z )= SL(2,\Z )/\Z_2$
(the quotient is over the global sign of transformations) \cite{cft}.
There are two generators, $T\ :\ \tau\to\tau + 1$
and $S\ :\ \tau\to -1/\tau$, satisfying the relations
$S^2=\left(ST\right)^3 =C$, where $C$ is the charge conjugation matrix,
$C^2=1$ \cite{cft}.

The invariance of the partition function under modular transformations
is therefore given by:
\be
Z_{\rm annulus}\left(\frac{-1}{\t},\frac{-\z}{\t} \right) = 
Z_{\rm annulus}\left(\t+1,\z \right) = Z_{\rm annulus}\left(\t,\z\right)\ .
\label{minv-zed}
\ee

Let us briefly recall from \cite{cz} the physical conditions corresponding to 
these symmetries.
The $S$ invariance amounts to a completeness condition for the spectrum of the
RCFT: upon exchanging time and space, it roughly imposes that the set of
states at a given time is the same as that ensuring time propagation
\cite{cft}. The $S$ transformation acts by a unitary
linear transformation in the finite basis of the characters (\ref{rcftz}):
\be  
\theta_a\left(-{1\over\tau},-{\zeta\over\tau}\right)=e^{i\varphi}
\sum_{b=1}^N\ S_{ab}\ \theta_b\left(\t,\z\right) \ ,
\label{linear-s}
\ee 
($\varphi$ is an overall phase and $N=p$ for Laughlin states).
The matrix $S_{ab}$ determines the fusion rules of the RCFT through
the Verlinde formula \cite{cft}; moreover, the dimension $N$ of the matrix
is equal to the Wen topological order of the Hall fluid \cite{wen}.

The action of the $T^2$ transformation is,
\be 
T^2:\qquad
Z\left(\tau +2,\zeta\right) \equiv {\rm Tr}\ 
\left[\cdots {\rm e}^{\ i2\pi\ 2\left(L^L_0 -L^R_0\right)} \right] = 
Z\left(\tau, \zeta\right)\ ,
\label{tcond}
\ee
namely it allows states with integer or half-integer spin on the
whole system.
These are electron-like excitations that carry the electric current
in and out of the system: thus, anyon excitations present 
on both edges, should combine between them to form global fermionic states.  
Note that the presence of fermions in the QHE implies 
the weaker invariance under $T^2$ rather than $T$: 
actually, $S$ and $T^2$ generate a subgroup of the modular
group, $\G_\theta \subset \G$ \cite{cft}.

The partition function should also be invariant under the 
transformations of the $\z$ variable, $\z\to\z+1$ and $\z\to \z+\t$, 
that can be geometrically interpreted as 
a coordinate on the torus.  In physical terms, these
correspond to conditions on the charge spectrum.
The first one,
\be
U:\qquad Z\left(\tau,\zeta +1\right) \equiv
{\rm Tr}\ \left[\cdots {\rm e}^{\ i2\pi\left(Q^R +Q^L\right)}
\right] = Z\left(\tau, \zeta\right)\ ,
\label{u-cond}\ee
requires that excitations possess total integer charge,
$Q^L+Q^R \in \Z$ (no quasiparticles are present in the bulk). 
Thus, fractionally charged excitations
at one edge must pair with complementary ones on the other
boundary. Consider, for example, a system at $\nu=1/3$ and add one
electron to it: it can split into one pair of excitations with, 
$(Q^L,Q^R)=(1/3,2/3), (0,1), (2/3,1/3), (1,0)$. 
The different splittings are related one to another by tuning the 
electric potential $V_o$ in (\ref{hconf}). 

The potential at the edges can be varied by adding localized magnetic 
flux inside the annulus;
the addition of one flux quantum leads to a symmetry of the spectrum,
as first observed by Laughlin \cite{wen}.
This is actually realized by the transformation $\zeta\to\zeta +\tau$, 
corresponding to $V_o \to V_o+1/R$ (in our notations $e=c=\hbar=1$)
\cite{cz}.
The corresponding invariance of the partition function is:
\be
V:\qquad Z\left(\tau,\zeta +\tau\right) = Z\left(\tau, \zeta\right)\ .
\label{vcond}\ee
This transformation is called ``spectral flow'', because
it amounts to a drift of each state of the theory into another one. 

In the case of the Laughlin theory, the  $T^2,S,U,V$  transformations
of the characters $\theta_\l$ (\ref{thetaf}) are given by \cite{cz},
\ba
T^2:\ \theta_\lambda\left(\tau+2,\zeta\right)\ & = &
{\rm e}^{ i2\pi\left({\lambda^2\over p}-{1\over 12} \right) }
\theta_\lambda\left(\tau,\zeta\right)\ ,
\nonumber\\
S:\ \theta_\lambda\left(-{1\over\tau},-{\zeta\over\tau}\right) & = &
{\rm e}^{i{\pi\over p}\R {\zeta^2 \over \tau} }\frac{1}{\sqrt{p}}
\ \sum_{\lambda^\prime=0}^{p-1}\
{\rm e}^{ i2\pi{\lambda\lambda^\prime\over p} } \
\theta_{\lambda^\prime}\left(\tau,\zeta\right)\ ,
\label{chitr}\\
U:\ \theta_\lambda\left(\tau,\zeta+1\right)\ & = &
{\rm e}^{ i2\pi\lambda/ p } \
\theta_\lambda\left(\tau,\zeta\right)\ ,
\nonumber\\
V:\ \theta_\lambda\left(\tau,\zeta+\tau\right)\ & = &
{\rm e}^{ -i{2\pi\over p}\left(\R\zeta +\R{\tau\over 2} \right) }
\theta_{\lambda+1}\left(\tau,\zeta\right)\ .
\nonumber
\ea
These formulas show that the generalized characters $\theta_\l$ carry a
unitary representation of the modular group, which is projective
for $\zeta\neq 0$ (the composition law is verified up to a phase).

Note that the charge transported between the two edges by adding one flux
quantum in the center ($V$ transformation) is equal to the Hall conductivity:
indeed, $\theta_\lambda(\zeta+\tau)\propto\theta_{\lambda+1}(\zeta)$
corresponds to $\nu=1/p$.
This provides a method to determine the value of $\nu$ from
the partition function.

The corresponding sums of right $\u1$ representations
are given by $\ov\theta^c_{\mu}$ carrying charge $Q^R=\mu/p+{\Z}$.
Finally, the $U$ condition (\ref{u-cond}), applied to $Z_{\rm annulus}$ 
(\ref{rcftz}), requires that left and right charges obey:
$\l+\mu=0 $ mod $p$. This form of the partition function 
also satisfies the other conditions, $T^2,S,V$,  by unitarity. 

Finally, the modular invariant partition function of Laughlin's states is:
\be
Z_{\rm annulus}=\sum_{\lambda=1}^p\ \theta_\l\ \ov\theta_\l \ .
\label{zedone}
\ee

The partition function for the disk geometry is obtained from that of the
annulus (\ref{zedone}) by letting the inner radius to vanish, $ R_L\to 0$ (see
Fig. \ref{z-geom}(b)).  To this effect, the variable $\ov{\t}$ in
$\ov{\theta}_\l$ should be taken independent of $\t$: $\I \t \neq - \I \ov\t$,
$v_R/R_R\neq v_L/R_L$. The annulus partition function is no longer a real
positive quantity but remains modular invariant, up to a global phase.
In the limit $ R_L\to 0$, the $\ov{\theta}_\l$ are dominated by their  
$|q|\to 0$ behavior: therefore, only the 
ground state sector remains in (\ref{zedone}), leading
to $\ov{\theta}_\l\to \d_{\l,0}$, up to zero-point energy contributions. 
One finds: $Z_{\rm disk}^{(0)}=\theta_0 $. 
In presence of quasiparticles in the bulk of the disk, with charge 
$Q_{\rm Bulk}=-a/q$, the condition of total integer charge selects 
another sector, leading to:
\be
Z_{\rm disk}^{(a)}=\theta_a \ .
\label{z-disk} 
\ee 
This is the desired result for the partition function on the disc geometry;
there are $N$ such functions, $a=1,\dots,N$, 
that transforms unitarily among themselves under
$S$ as a $N$-dimensional vector. 
Note here a manifestation of the general correspondence between bulk
and edge excitations, that can be proven by using Chern-Simons theory
but is actually valid for general RCFTs \cite{jones}: bulk excitations are
equivalent, as much as the low-energy theory is concerned, to an
edge with radius shrinking to zero; in this limit, the tower of excitations
in that sector decouples.

\subsection{Physical conditions for the spectrum}

The construction of RCFTs for quantum Hall states, both
Abelian and non-Abelian, has been relying on a set of conditions
for the spectrum of charge and statistics that implement
the properties of electron excitations \cite{hiera}; 
they should have:

A) integer charge;

B) Abelian fusion rules with all excitations;
 
C) fermionic statistics among themselves (half-integer spin);

D) integer statistics with all other excitations (integer exponent of mutual
exchange).

The (B) and (D) conditions characterize the operator-product
expansion of the conformal field $\Phi_e(z)$, representing the electron,
with the field $\Phi_i(w)$ of a quasiparticle: for $z\to w$, this is,
\be
\Phi_e(z)\ \Phi_i(w)\sim \left(z-w \right)^{h_{ei}}\ \Phi_{e \times i}(w)\ .
\label{ope-e}
\ee
The mutual statistics exponents is given by the conformal dimensions,
\be
h_{ei}=-h_e - h_i + h_{e\times i}\ ,
\label{h-stat}
\ee
respectively of: the electron field, the $i$-th quasiparticle and their fusion
product, $\Phi_{e\times i}=\Phi_e \times \Phi_i$.
In general, excitations  are made 
of a neutral part, described by a non-trivial RCFT, 
typically a coset theory $G/H$,
and by a charged
part (Luttinger liquid, i.e. a $\u1$ RCFT) \cite{cft}: the fields
in the above formulas are made of neutral and charged parts, and
their dimensions $h$ have contributions from both of them.
The requirement of integer statistics of the electron with all excitations,
$h_{ei}\in \Z$, is motivated by the properties of many-body wave functions
(describing, e.g.  states with a quasiparticle of $i$-th type),
that are described by correlators of the same RCFT for edge excitations.
This bulk-edge correspondence follows again from the 
description of Hall fluids in terms of the Chern-Simons theory
and is believed to be true for all rational CFTs \cite{jones}.

The condition of Abelian fusion rules restricts to one the number of
terms in the r.h.s. of the operator product expansion (\ref{ope-e}):
indeed, if there were more terms, all the corresponding $h_{ei}$ 
exponents would need to be simultaneously integer, 
a condition generically impossible to achieve.
Even if it were satisfied, this would lead to a degeneracy of $n$-electron
wave functions and to a unacceptable degenerate ground state.
Non-Abelian fusion rules and associated degeneracies
(the quantum dimensions $d_a$) are possible for quasiparticles 
but not for electrons (within the RCFT description, at least \cite{cz2}).

In the following, we show that
the electron conditions (A)-(D) can be re-derived from the requirement of
modular invariance of the partition function.
We find that, upon choosing the RCFT for the neutral part 
of excitations and identifying the field representing the electron, 
the modular conditions are sufficient to self-consistently determine 
the charge and statistics spectrum, as well as the filling 
fraction\footnote[1]{
And the spin parts, for non-polarized Hall fluids.},
without the need of additional physical hypotheses.

Let us compare (A)-(D) with the modular conditions introduced
in the previous section for Laughlin states.
The condition (A) is clearly the same as the $U$ modular invariance
(\ref{u-cond}), whose solution is the extended character $\theta_\l$ 
(\ref{rcftz}), that resums electron excitations added to
the $\l$ quasiparticle.

The condition (B) of Abelian fusion rules of the electron has a natural 
correspondent in RCFT, where a field with such property
is called a ``simple current'' $J \ (\equiv\Phi_e )$. 
The notion of simple current was 
introduced for orbifold theories and their modular invariant
partition functions, as we now briefly recall \cite{cft}\cite{fss}.

The action of the simple current is indicated by
$J\times \F_i =\F_{J(i)} (\equiv \F_{e \times i})$;
it implies an Abelian discrete symmetry
in the theory that is generated by $\exp(2i\pi Q_J)$, with:
\be
Q_J\left(\F_i \right) = h_J + h_i - h_{J(i)} \qquad {\rm mod}\ 1 .
\label{q-def}
\ee
This charge is the exponent for the monodromy discussed in (\ref{h-stat})
 and is conserved in the fusion rules.
The fields $\F_i$ can be organized in orbits, each orbit containing the 
fields generated by the repeated fusion with the simple current.
The simple current and its powers generate an Abelian
group by fusion that is called the ``center'' ${\cal G}$ of the 
conformal field theory.

The modular invariant partition function can be obtained by the
orbifold construction corresponding to 
modding out the symmetry associated to the simple current.
The result has the general form \cite{fss}:
\be
Z = \sum_{{\rm orbits}\ a  |\ Q_J(a) = 0}\ \left|{\cal S}_a\right| \ 
\left | \sum_{J \in {\cal G}/{\cal S}_a} \chi_{J(i_a)} \right |^2 \ 
= \sum_a \left | \theta_a \right |^2;
\label{ext-z}
\ee
in this equation, $a$ labels the orbits, $ i_a$ 
is a representative point on each orbit, and  $|{\cal S}_a|$ is the order of
the stabilizer ${\cal S}_a$ of the orbit $a$, i.e. the subgroup
of ${\cal G}$ acting trivially on any element $i$ in $a$.
The proof of the general expression (\ref{ext-z}) is not trivial 
and it involves the symmetry of the $S$ matrix under the $J$ action:
$S_{i,J(k)}=S_{i,k}\exp(2i\pi Q_J(i))$.
The modular invariants (\ref{ext-z}) can be considered as diagonal
 invariants with respect to the basis of the extended chiral algebra,
whose characters are $\theta_a$. 
In the QHE case, the stabilizer is trivial and the $J$ action
has no fixed points, owing to the additivity of the
physical charge carried  by the electron \cite{kt}.
We recognize in (\ref{q-def}),(\ref{ext-z}), the (D) condition 
(\ref{h-stat}) derived from QHE wave functions.
Note that in the Abelian case this condition follows immediately 
from $U$ invariance \cite{cvz}.

Let us further analyze the solution (\ref{ext-z}); the
$T^2$ invariance of the extended characters for the 
ground state and $a$-th quasiparticles, respectively 
$\theta_0$ and $\ov{\theta}_a$, implies:
\be
2\ h_e= M\ ,\qquad\quad 2\ h_{e\times i} -2\ h_i = N\ , \qquad\qquad
M,N \ {\rm integers.}
\label{t2-cond}
\ee
The condition (C) of half-integer electron spin requires $M$ to be odd,
i.e the even case is excluded for physical reasons (although sometimes
allowed for bosonic fluids).
Therefore, we should consider algebra extensions by  half-integer
spin currents, as in the case of the Neveu-Schwarz sector 
of supersymmetric theories \cite{cft}. 
(Note that the integer $N$ in (\ref{t2-cond})
is also odd by the (D) condition).

In conclusion, we have re-derived the standard physical conditions (A)-(D) 
on Hall excitations from modular invariance of RCFT partition functions.
They have been found to be diagonal invariants for extended symmetry
algebras that are obtained in the orbifold construction 
through simple currents (in the present case without fixed points) 
\cite{fss}.

In the following analysis of non-Abelian Hall fluids, we shall find that the
(A)-(D) conditions (i.e. modular invariance) straightforwardly determine the
charge and statistics spectrum of excitations, for a given choice of 
neutral RCFT and electron field (simple current). These results are relevant
for model building: in the literature, the derivation of the theory pertaining
to a given plateau often involves a combination of technical arguments and
physical arguing, that might suggest a certain degree of arbitrariness
in the construction of the theory, which is however not present.

We remark that the neutral RCFT may possess more than one simple current that
could be used as electron field, although a preferred choice may exist,
e.g. the lowest-dimensional field.  This cannot be considered as an
ambiguity of the construction, because the choice of electron field is part of
the definition of the theory: different electrons correspond to different Hall
states, with different charge spectra, filling fraction etc, all quantities
being determined self-consistently.

Another possibility for RCFTs with two simple currents
is that of using both of them simultaneously for building a modular invariant
with further extended symmetry. This issue will be discussed in the
conclusions.

\subsection{Example: Read-Rezayi states}

In the following, the conditions (A)-(D) will be illustrated by
rederiving the spectrum and partition functions \cite{cgt2} \cite{cvz} 
of Read-Rezayi states \cite{rr} with filling fractions:
\be
\nu =2 + \frac{k}{kM+2}\ , \qquad k=2,3,\dots, \quad M=1,3,\dots
\label{nu-RR}
\ee
The Read-Rezayi theory is based on the neutral $\Z_k$ para\-fermion theory
(${\rm PF}_k$) with central charge $c=2(k-1)/(k+2)$, that can be
described by the standard coset construction 
${\rm  PF}_k=\wh{SU(2)}_k/\u1_{2k}$ \cite{qiu}.  
Form the coset, we find that neutral sectors are characterized by
a pair quantum numbers for the representations of the algebras
in the numerator and denominator: these are $(\ell, m)$, equal to twice 
the $SU(2)$ spin
and spin component, respectively ($m=\ell$ mod $2$).

\begin{figure}[t]
\begin{center}
\includegraphics[width=5cm]{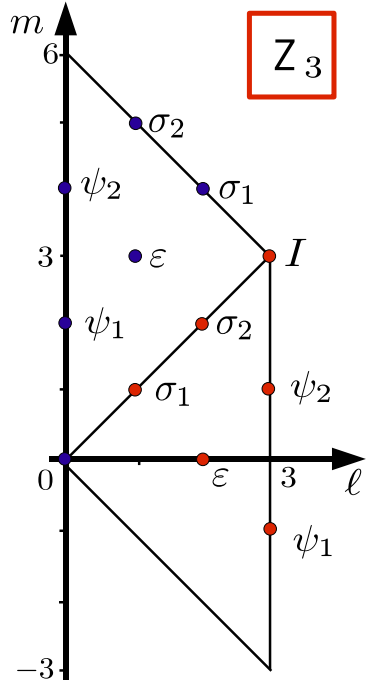}
\end{center}
\caption{Diagram of $\Z_3$ parafermion fields with field symbols}
\label{pf-fig}
\end{figure}

The dimensions of parafermionic fields $\f^\ell_m$ are given by:
\ba
h_m^\ell& =&\frac{\ell(\ell+2)}{4(k+2)}-\frac{m^2}{4k}\ ,
\nl
&& \ell=0,1,\dots,k, \qquad -\ell <m\le \ell, \quad \ell=m
{\rm \ mod\ }2\ .
\label{hlm}
\ea
The $\Z_3$ parafermion fields are shown in Fig. \ref{pf-fig}:
the coset construction implies that the $m$ charge is defined modulo
$2k$ \cite{qiu}; indeed, the fields are repeated once outside the fundamental 
$(\ell,m)$ domain (\ref{hlm}) by the reflection-translation,
$(\ell,m)\to (k-\ell,m+ k)$,
\be
\f^\ell_m =\f^{k-\ell}_{m-k}\ , \qquad \ell=0,1,\dots,k,\qquad
l<m\le 2k-l\ ,
\label{refl}
\ee
also called ``field identification'' \cite{gepner}.

The fusion rules are given by the addition of the
$\wh{SU(2)}_k$ spin and $\u1_{2k}$ charge:
\be
\f_m^\ell\cdot\f_{m'}^{\ell'} =
\sum_{\ell''=|\ell-\ell'|}^{{\rm min}(\ell+\ell',2k-\ell-\ell')}\ 
\f^{\ell''}_{m+m'\ {\rm mod}\ 2k}\ .
\label{fus-RR}
\ee
The fields in the theory are called: parafermions, $\psi_j =\f_{2j}^0$,
$j=1,\dots,k-1$; spin fields, $\s_i=\f_i^i$ ; and other fields. The
parafermions have Abelian fusion rules with all the fields; among themselves,
these are: $\psi_i\times\psi_j=\psi_n$, with $n=i+j$ mod $k$.
The basic parafermion $\psi_1$ represents the neutral component of the
electron in the Read-Rezayi states: the fusion $\left(\psi_1\right)^k = I$,
describes the characteristic clustering of $k$ electrons in the ground
state-wave function \cite{rr}.

The excitations of the full theory are found by attaching 
$\u1$ vertex operators to the parafermion fields: for the electron and
a generic quasiparticle, we write,
\be
\Phi_e=e^{i\a_0\vf}\ \psi_1\ , \qquad \Phi_i = e^{i\a\vf}\ \f_m^\ell\ ,
\label{rr-fields}
\ee 
with triplets of quantum numbers $(\a_0,0,2)$ and $(\a,\ell,m)$,
respectively.
The mutual statistics exponent (\ref{h-stat}) is,
\be
h_{ei}=\frac{(\a+\a_0)^2-\a^2-\a_0^2}{2}+
h^\ell_{m+2}- h^\ell_{m} - h^0_{2} \ .
\label{h-pf}
\ee
Upon substituting (\ref{hlm}), the conditions (D) reads:
\be
(D)\ :\ \a \a_0 -\frac{m}{k} = N , \qquad \quad {\rm integer}.
\label{cond-pf}
\ee
In particular, for the electron with itself, we find
$\a_0^2 -2/k =M$ integer; in 
combination with the condition (C) of half-integer electron spin,
\be
(C)\ :\ 2h_e=2h^0_2 + \a_0^2 = 2+ M  ,\qquad {\rm odd \ integer},
\label{cond-el}
\ee 
it determines,
\be
\a^2_0 =\frac{2+kM}{k} ,\qquad M \ {\rm odd \ integer}.
\ee

The electric charge $Q$ of excitations is proportional to the $\u1$ charge, 
$Q=\mu \a$: the constant $\mu$ is fixed by assigning $Q=1$ to the electron,
i.e. $\mu=1/\a_0$ ((A) condition).

In conclusion, the (A)-(D) conditions determine the charge and 
spin (i.e. half statistics) of excitations
in the Read-Rezayi theory as follows:
\ba
Q &=& \frac{\a\a_0}{\a_0^2} =\frac{Nk+m}{2+Mk}=\frac{q}{2+Mk}\ ,
\label{spec-q}\\
J &=&h_m^\ell + \frac{1}{2}Q^2\a_0^2 = h_m^\ell+\frac{q^2}{2k(2+kM)}\ .
\label{spec-h}
\ea
The excitations are characterized by triplets of integer labels 
$(q,\ell,m)$, where $q$ is the charge index.
From the Abelian part of conformal dimensions (\ref{spec-h}), we find that
the Luttinger field is compactified into $p$ sectors (cf. (\ref{thetaf})),
$p=k(2+kM)$, and is indicated by $\u1_p$; 
the charge index $q$ is thus defined modulo $p$.  
We write $p=k\hat{p}$, where $\hat{p}=2+kM$ is the denominator of the
fractional charge; moreover, equation (\ref{spec-q}) shows that $q$ is coupled
to the $SU(2)$ spin component $m$ by the selection rule:
\be
q = m \quad {\rm mod\ } k\ , \qquad\quad 
\left( q \quad {\rm mod\ } \hat{p}=(kM+2) \ , 
\quad m \quad {\rm mod\ } 2k\right) .
\label{pf-rule}
\ee

We thus recover the $\Z_k$ ``parity rule''  of  Ref. \cite{cgt2}
that constraints charge and neutral quantum numbers of Read-Rezayi
quasiparticle excitations. Earlier derivations of this rule 
were based on some physical conditions,
such as a relation with a ``parent Abelian state'',
that were useful as motivations but actually not
necessary (see, however, section 4.1.1).  The present derivation shows that the
relevant information is the type of RCFT for neutral excitations and the
identification in this theory of the field $\psi_1$ representing 
the electron.

As outlined in the previous section, the derivation of  
annulus partition functions requires the solution of the modular invariance
conditions: in particular, the $U$ symmetry (\ref{u-cond}) requires to put
each basic anyon together with its electron excitations, leading to 
sectors of charge $Q=\l/p +\Z$ described by the RCFT extended character
$\theta_\l$ as in (\ref{thetaf}). The addition of an electron 
changes the integer labels of excitations as follows:
\be
\left(q,m,\ell \right) \ \to \ \left(q+\hat{p},m+2,\ell \right)\ .
\label{e-add}
\ee

Therefore, the extended characters $\theta_\l$ of the theory 
$PF_k \otimes \u1_p$ are made of products of characters 
of the charged and neutral parts, whose indices obey the parity rule 
(\ref{pf-rule}) and are summed over according to (\ref{e-add}).
The charged characters are given by the functions $K_q(\t,k\z;k\hat p)$
introduced earlier in (\ref{thetaf}), with parameters chosen 
to fit the fractional charge and the Abelian conformal dimension 
in (\ref{spec-q}).
The $\Z_k$ parafermionic characters are denoted by $\c^\ell_m(\t;2k)$ and have
rather involved expressions; actually it is enough to know
their periodicities,
\ba
\c^\ell_m&=&\c^\ell_{m+2k}=\c^{k-\ell}_{m+k} \ , \quad
m=\ell\ {\rm mod}\ 2,
\nl
\c^\ell_m&=&0\, \qquad\qquad\quad\qquad m=\ell+1\ {\rm mod}\ 2\ ,
\label{pf-char}
\ea
and modular transformation:
\ba
&&\c^\ell_m\left(-1/\t;2k\right) 
= \frac{1}{\sqrt{2k}} \sum_{\ell'=0}^k \ \sum_{m=1}^{2k}\
e^{-i2\pi\frac{ mm'}{2k}}\ s_{\ell,\ell'}
\ \c^{\ell'}_{m'}(\t;2k)\ ,
\nl
&& s_{\ell,\ell'}
= \sqrt{\frac{2}{k+2}} \ 
\sin\left(\frac{\pi (\ell+1)(\ell'+1)}{k+2}  \right)\  .
\label{pf-s}
\ea
As recalled in Appendix A, this transformation can be obtained from 
the coset construction $PF_k=\wh{SU(2)}_k/\u1_{2k}$ \cite{qiu}.

The extended characters are thus given by  products
$K_q \c^\ell_m$, with $q=m $ mod $k$ and $\ell=m$ mod $2$:
adding one electron to the earlier product
gives the term $K_{q+\hat{p}} \c^\ell_{m+2}$; by
continuing to add electrons until a periodicity is found, 
one obtain the expressions \footnote[2]{
Hereafter, we disregard the non-analytic prefactor of $K$ 
for ease of presentation.},
\ba
\theta_a^\ell &=&\sum_{b=1}^k K_{a+b\hat p}(\t,k\z;k\hat p)\ 
\c^\ell_{a+2b}(\t;2k) \ , 
\nl
&&\begin{array}{l}
a=0,1,\dots, \hat p-1,\ \hat p=kM+2,\\
\ell=0,1,\dots,k,\\
a = \ell\ {\rm mod}\ 2,
\end{array}
\label{pf-th}
\ea
that reproduce the charge and statistics spectrum (\ref{spec-q}),
(\ref{spec-h}).

The expression (\ref{pf-th}) corresponds to the following solution of
the parity rule (\ref{pf-rule}):
\ba
&& q=a+b\hat{p}, \qquad a=1, \dots, \hat{p}, \qquad b=1,\dots, k\ ,
\nl
&& m=a+2b\ ,
\label{q-decomp}
\ea
(note $\hat p=2$ mod $k$). The other solution $m=k+a+2b$ mod $2k$ would lead
to the same expressions with shifted index, $\theta^{k-\ell}_a$, owing to
the field identification of parafermion fields
$(\ell,m)\sim (k-\ell, m \pm k)$.

The $\theta_a^\ell$ characters have the periodicity, 
$\theta^\ell_{a+\hat p}=\theta^{k-\ell}_a$, that
explains the ranges of indices indicated in (\ref{pf-th}); the dimension of 
the basis of $\theta^\ell_a$ characters is therefore given 
by $\hat{p}(k+1)/2$, in agreement with the value of the topological order 
of Read-Rezayi states \cite{rr} \cite{cgt2}.
Moreover, the $V$ transformation of these characters reads,
$\theta_a^\ell(\z +\t)\sim \theta(\z)_{a+k}^{k-\ell}$,
showing that the charge $\Delta Q=k/\hat p$ is created by adding one flux
quantum: we thus recover the value of the filling fraction $\nu=k/(Mk+2)$ 
with $M$ odd.

The final step is to find the modular transformations of $\theta_a^\ell$,
that is given by \cite{cvz} (see Appendix A):
\be
\theta^\ell_a(-1/\t) = \d^{(2)}_{a,\ell}\ 
\frac{1}{\sqrt{\hat p}}
\sum_{a'=1}^{\hat p}\sum_{\ell'=0}^k\ 
e^{-i2\pi \frac{aa' M}{2\hat p}}\ s_{\ell,\ell'} \ 
\theta^{\ell'}_{a'}(\t) ,
\label{pf-th-s}
\ee
where the delta modulo two tells that $ \theta^\ell_a(-1/\t) $ vanishes for
$a=\ell+1$ mod $2$ (we also disregard the global phase $\propto\R(\z^2/\t)$
acquired by the characters).  In Appendix A, the check of the unitarity of the
$S$ matrix confirms that the extended characters $\theta_a^\ell$
form an independent basis.

We remark that the coupling of neutral and charged parts by the $\Z_k$ parity
rule and the sum over electron excitations amounts to a projection in
the full $K_\l \c^\ell_m$ tensor space to a subspace of dimension 
$1/k^2$ smaller: this reduction by a square factor is a standard property 
of $S$ transformations (i.e.  discrete Fourier transforms).
We also note that the $S$ matrix is factorized into charged and neutral
parts, where the latter is the naive expression for the $\wh{SU(2)}_k$ part,
although the sectors are not factorized at all, as shown by 
the extended characters $\theta_a^\ell$. 

Finally, the annulus partition function of Read-Rezayi states is given by 
the diagonal sesquilinear form,
\be
Z_{\rm annulus}^{\rm RR} =\sum_{\ell=0}^k\ \ 
\sum_{a=0 \atop a=\ell \ {\rm mod}\ 2}^{\hat p-1}
\ \left\vert\ \theta^\ell_a\ \right\vert^2\ ,
\label{pf-zann}
\ee
that solves the $(S,T^2,U,V)$ conditions of section 2.2. 

For example, the expression of the $k=2$ Pfaffian state is
as follows. The $\Z_2$ parafermions
are the three fields of the Ising model:
$\f^0_0=\f^2_2=I$, $\f^1_1=\f^1_3=\s$ and $\f^0_2=\f^2_0=\psi$,
of dimensions $h=0,1/16,1/2$, respectively.
For $\nu=5/2$, i.e. $M=1$ in (\ref{nu-RR}), the Pfaffian theory possesses
$6$ sectors. The partition function is: 
\ba
Z_{\rm annulus}^{\rm Pfaffian} &=&
\left\vert  K_0 I+ K_4\psi\right\vert^2\ +
\left\vert  K_0\psi+K_4 I \right\vert^2\ +
\left\vert \left( K_1+ K_{-3}\right)\s\right\vert^2
\nl
&+&
\left\vert K_2 I +K_{-2}\psi \right\vert^2\ +
\left\vert  K_2\psi+K_{-2} I \right\vert^2\ +
\left\vert \left( K_3+ K_{-1}\right)\s\right\vert^2\ .
\label{ising-zann}
\ea
where the neutral characters are written with the same symbol of the field
and the charged ones are, $K_\l=K_\l(\t,2\z;8)$, $K_\l=K_{\l+8}$, 
with charge $Q=\l/4+2\Z$.
The first square term in $Z$ describes the ground state and its electron
excitations, such as those in $K_4\psi$ with $Q=1+2\Z$; in the third
and sixth terms, the characters $K_{\pm 1}\s$ contain the basic
quasiparticles with charge, $Q=\pm 1/4$, and non-Abelian fusion rules
$\s\cdot\s=I+\psi$.  The other three sectors are less familiar: the second
one contains a neutral Ising-fermion excitation (in $K_0\psi$) and the
4th and 5th sectors describe $Q=\pm1/2$ Abelian quasiparticles.

As in section 2.1, the partition function on the disk is given by
the extended character $\theta_a^\ell$, with indices selected by the
quasiparticle type in the bulk; if there are many of them, their indices are 
combined by using the fusion rules to find the edge sector $(a,\ell)$.

In conclusion, the annulus and disk partition functions completely determine
the Hilbert space of edge excitations and the fusion rules through
the Verlinde formula. Physical applications to current experiments
will be describe in section 4.

\section{Partition functions of non-Abelian Hall states}

In this section we obtain the partition functions of other
non-Abelian states that have been proposed to describe plateaus
with $2<\nu<3$: the Wen non-Abelian fluids (NAF) \cite{bw}, 
the anti-Read-Rezayi states ($\ov {\rm RR}$) \cite{antiRR},
the Bonderson-Slingerland hierarchy (BS) \cite{bs}, and
finally, the non-Abelian spin-singlet
state (NASS) \cite{nass}\cite{ardonne}. 
All these states have been considered as phenomenologically
interesting in the recent literature searching for signatures of 
non-Abelian statistics in the quantum Hall effect.

From the technical point of view, non-Abelian states can be built out of the
$\u1_p$ charged part and a neutral part given by any RCFT that possess at
least one ``simple current'' \cite{cft}, a field with Abelian fusion rules with
all the others that can be associated to the electron excitation.
However, only a limited number of such constructions have received support 
by $(2+1)$-dimensional microscopic physics, that is based 
on wave-functions and analytic/numerical study of spectra searching for
corresponding incompressible states.

\subsection{$SU(2)$ non-Abelian fluids}

Some time ago, Block and Wen \cite{bw} considered the natural choice
of RCFT with affine symmetry $\wh{SU(m)}_k$ and the associated $SU(m)$ 
non-Abelian Chern-Simons theory, starting from the physical idea of 
breaking the electron excitation into $k$ fermions called ``partons''.
In such theories, there always are one or more simple currents. 
We shall limit ourselves to the simplest $SU(2)$ case, that has been
recently considered in relation with the physics of second Landau level
\cite{na-interf}\cite{dopp} 
and also serves as a starting point for other non-Abelian fluids.

We shall obtain the 
annulus partition function for the RCFT $\wh{SU(2)}_k\otimes\u1_p$.
The  $\wh{SU(2)}_k$ theory is characterized by the primary fields
$\phi^\ell$, for $\ell=0,1,\dots,k$, with dimensions
$h_\ell=\ell(\ell+2)/(4(k+2))$. The simple current is $\phi^k$
with $h_k=k/4$ and fusion rules given by (cf. (\ref{fus-RR})): 
\be
\phi^k \times \phi^\ell =\phi^{k-\ell}\ , \qquad 0\le \ell\le k\ ,
\label{su2-sc}
\ee
which realizes a $\Z_2$ parity among the neutral sectors.
Following the steps outlined in section 2.2, we introduce the fields 
$\Phi_e=\phi^k \exp(i\a_0\vf)$, $\Phi_i=\phi^\ell \exp(i\a\vf)$
corresponding to the electron and to a quasiparticle excitation, respectively.

The condition (D) on the mutual statistics exponent (\ref{h-stat}) 
of these excitations and of the electron with itself imply, respectively:
\be
\a\a_0=\frac{2N + \ell}{2} \ ,\qquad\quad 
\a_0^2=\frac{2M + k}{2}\ ,\qquad N,M\in \Z.
\label{su2-stat}
\ee
The charge and spin of excitations are therefore:
\ba
Q &=& \frac{\a}{\a_0} =\frac{2N+\ell}{2M+k}=\frac{q}{2M+k}\ , \qquad\quad
M+k\ \ {\rm odd\ integer},
\nl
J &=&h^\ell + \frac{1}{2}Q^2\a_0^2 = 
\frac{\ell(\ell+2)}{4(k+2)}+\frac{q^2}{4(2M+k)}\ .
\label{su2-qh}
\ea
In particular, the condition (C) of half-integer electron spin 
requires that $(M+k)$ is odd.

The $\u1_p$ contribution to the $J$ spectrum identifies the 
compactification parameter $p=2\hat{p}$, with
$\hat{p}=(2M+k)$ the fractional-charge denominator.
A quasiparticle is characterized by the integer pair $(\ell,2N+\ell)$
of $SU(2)$ spin and charge, that are constrained
by the parity rule: $q=\ell$ mod $2$.
The additions of one (two) electrons to a quasiparticle
cause the following shifts (i.e. changes of sector):
\be
(\ell,2N+\ell)\ \to\ (k-\ell,2N+\ell+\hat{p})
\ \to\ (\ell,2N+\ell+2\hat{p})\ ,
\label{su2-shifts}
\ee
that involves two non-Abelian sectors only.

Therefore, we are lead to consider extended characters $\theta^\ell_a$
of the full theory, labeled by spin $\ell$ and charge $a$ indices,
that are made of products of charged characters, $K_q=K_q(\t,2\z;2\hat{p})$
in the notation of (\ref{thetaf}), and $\wh{SU(2)}_k$ characters
$\c^\ell(\t)$, as follows:
\be
\theta^\ell_a=K_a\ \c^\ell +K_{a+\hat{p}}\ \c^{k-\ell}\ , \qquad\qquad
\begin{array}{l}
a=0,1,\dots, \hat p-1,\ \hat p=2M+k,\\
\ell=0,1,\dots,k,\\
a = \ell\ {\rm mod}\ 2.
\end{array}
\label{su2-theta}
\ee
Note that only two terms are needed in the sums, owing to the 
mentioned $\Z_2$ symmetry.
As explained in section 2.2, the topological order of the $SU(2)$ NAF
is given by the number of independent $\theta^\ell_a$ characters:
the symmetry $\theta_{a+\hat{p}}^{k-\ell}=\theta_a^\ell$ confirms
the index ranges indicated in (\ref{su2-theta}).
Furthermore, the filling fraction can be obtained from 
the $V$ transformation of $K_a$ characters: in summary, 
the $SU(2)$ NAF fluids are characterized by the values,
\be
T.O.=(k+1)\frac{2M+k}{2}\ , \qquad
\nu=\frac{2}{2M+k}\ ,\qquad M+k\ {\rm odd}.
\label{su2-nu}
\ee

The modular transformations of extended characters can be obtained by those
of the components $K_a$ and $\c^\ell$ introduced in earlier sections:
the result is (see Appendix A): 
\be
\theta^\ell_a(-1/\t) = 
\frac{1}{\sqrt{\hat p}}
\sum_{a'=0}^{\hat {p}-1}\sum_{\ell'=0}^k\ \d^{(2)}_{a,\ell}\ 
e^{i2\pi \frac{aa'}{2\hat p}}\  s_{\ell,\ell'}\ \d^{(2)}_{a',\ell'}\ 
\theta^{\ell'}_{a'}(\t) ,
\label{su2-s}
\ee
where $s_{\ell,\ell'}$ is the $\wh{SU(2)}_k$ $S$-matrix (\ref{pf-s}).
The $S$-matrix of the full theory is again factorized in Abelian
and non-Abelian parts (up to details) and is unitary.

In conclusion, the annulus partition function is given by the
diagonal combination of extended characters,
\be
Z_{\rm annulus}^{\rm NAF} =\sum_{\ell=0}^k\ \ 
\sum_{a=0 \atop a=\ell \ {\rm mod}\ 2}^{\hat p-1}
\ \left\vert\ \theta^\ell_a\ \right\vert^2\ ,
\label{naf-zann}
\ee
with ranges of parameters given by (\ref{su2-theta}).
Each individual $\theta^\ell_a$ is the partition function on the disk geometry
in presence of a specific bulk excitation.

\subsection{Anti-Read-Rezayi fluids}

It has been recently proposed \cite{antiRR} that a particle-hole conjugate of
the $M=1$ Read-Rezayi fluids could be realizable in the second Landau level,
with filling fractions $\nu-2=1- k/(k+2)=2/(k+2)$.
In particular, for $\nu=5/2$ the Pfaffian state may compete with 
its conjugate state.
The fluid of RR holes inside the $\nu=3$ droplet possesses an additional
edge, leading to the CFT $\ov{SU(2)_k/U(1)_{2k}\otimes U(1)_p}\otimes \u1$.
The Luttinger liquids on edges of opposite chirality interact through
impurities and re-equilibrate: the result of this process
was shown to lead to the $\ov{SU(2)_k}\otimes\u1_p$ edge theory \cite{antiRR}
for the so-called anti-Read-Rezayi state ($\ov{\rm RR}$).

The partition function of this theory can be obtained as in the NAF case of
the previous section, with little modifications due to the different chirality
of the neutral sector.  The electron and quasiparticle fields are given by
$\Phi_e=\ov{\phi^k} \exp(i\a_0\vf)$ and $\Phi_i=\ov{\phi^\ell} \exp(i\a\vf)$,
respectively (note that $\phi^k$ is the unique simple current of the
$\wh{SU(2)_k}$ theory).

In the mutual statistics exponent, the chiral and antichiral conformal
dimensions should be subtracted leading to the conditions, 
\be 
\a\a_0=\frac{2N - \ell}{2} \ ,\qquad\quad \a_0^2=\frac{2M + k}{2}\ ,
\qquad N,M\in \Z.
\label{arr-stat}
\ee
The charge and spin of excitations are therefore:
\ba
Q &=& \frac{\a}{\a_0} =\frac{2N-\ell}{2M+k}=\frac{q}{2M+k}\ , \qquad\quad
M\ \ {\rm odd}
\nl
J &=&- h^\ell + \frac{1}{2}Q^2\a_0^2 = 
- \frac{\ell(\ell+2)}{4(k+2)}+\frac{q^2}{4(2M+k)}\ .
\label{arr-qh}
\ea
In particular, the condition (C) of half-integer electron spin 
requires that $M$ is odd.

The $\u1_p$ compactification parameter is
$p=2\hat{p}$, where
$\hat{p}=(2M+k)$ is the denominator of the fractional charge.
A quasiparticle is characterized by the integer pair $(\ell,2N-\ell)$
of $SU(2)$ spin and charge, that are again constrained
by $q=\ell$ mod $2$.

Following the same steps of the NAF case (cf. (\ref{su2-shifts}),
(\ref{su2-theta})), we obtain the extended characters $\theta^\ell_a$,
\be
\theta^\ell_a=K_a\ \ov{\c^\ell} +K_{a+\hat{p}}\ \ov{\c^{k-\ell}}\ , \qquad\qquad
\begin{array}{l}
a=0,1,\dots, \hat p-1,\ \hat p=2M+k,\\
\ell=0,1,\dots,k,\\
a = \ell\ {\rm mod}\ 2,
\end{array}
\label{arr-theta}
\ee
that are products of charged characters $K_q=K_q(\t,2\z;2\hat{p})$, of same
period $p=2(2M+k)$ as in the NAF case, and  
conjugate $\wh{SU(2)}_k$ characters.

The values of topological order and filling fractions of 
$\ov{\rm RR}$ fluids are given by the NAF expressions (\ref{su2-nu}),
\be
T.O.=(k+1)\frac{2M+k}{2}\ , \qquad
\nu=\frac{2}{2M+k}\ ,\quad M\ {\rm odd};
\label{arr-nu}
\ee
only the parity of $M$ is different.
The $M=1$ case mentioned at the beginning is recovered.

The modular transformations of $\ov{\rm RR}$ (\ref{arr-theta}) 
are the same as in the NAF case (\ref{su2-s}), 
because the $\wh{SU(2)}_k$ $S$-matrix
is real and thus not affected by conjugation.
The annulus partition function is finally given by:
\be
Z_{\rm annulus}^{\ov{\rm RR}} =\sum_{\ell=0}^k\ \ 
\sum_{a=0 \atop a=\ell \ {\rm mod}\ 2}^{\hat p-1}
\ \left\vert\ \theta^\ell_a\ \right\vert^2\ .
\label{arr-zann}
\ee

\subsection{Bonderson-Slingerland hierarchy}

In Ref. \cite{bs}, the authors considered the realization of hierarchical Hall
states in the second Landau level, that are build over the Pfaffian 
$\nu=5/2$ state. 
In the Jain construction \cite{jain}, the wave functions are given by
$\Psi=\Delta^{2p} \c_n$, where $\c_n$ is relative to $n$ filled Landau levels
and $\Delta^{2p}$ is an even power of the Vandermonde factor, 
leading to the filling fraction $\nu=n/(2pn+1)$.
The (projected) Jain wave function, interpreted within the Haldane-Halperin 
hierarchical construction, can be transposed into the second Landau level,
and give rise to the Bonderson-Slingerland wave functions of the form
$\Psi={\rm Pf}(1/(z_i-z_j))\Delta^{M} \c_n$, with filling fractions
$\nu-2=n/(nM+1)$, with $M$ odd.

From earlier studies of edge excitations of Jain states \cite{hiera},
\cite{cz},
we know that the associated CFT has central charge $c=n$ and is 
Abelian with extended symmetry $\wh{SU(n)}_1\otimes \u1$:
this theory is described by a specific 
$n$-dimensional charge lattice that includes the $SU(n)$ root lattice. 
The Bonderson-Slingerland hierarchy is thus realized by the CFT
$\u1_p\otimes \wh{SU(n)}_1\otimes {\rm Ising} $, where the last two
factors are neutral. The respective conformal dimensions are \cite{cft}
\cite{cvz}:
\be
\begin{array}{lcll}
\u1_p &:& h_\a=\frac{\a^2}{2}\ , &
\\
\wh{SU(n)}_1 &:& h_\b=\frac{\b(n-\b)}{2n}\ , &\b=0,1,\dots, n-1,
\\
{\rm Ising} &:& h_m=0,\ \frac{1}{16},\ \frac{1}{2}\ , &m=0,1,2.
\end{array}
\label{bs-dim}
\ee
The electron and quasiparticle excitations are made by triplets of fields,
\be
\Phi_e=\r_1\; \phi^0_2 \exp(i\a_0\vf)\ , \qquad
\Phi_i=\r_\b\; \phi^\ell_m \exp(i\a\vf)\ .
\label{bs-fields}
\ee
In this equation, we indicated the $\wh{SU(n)}_1$ fields by 
$\r_\b$, with $\b$ mod $n$, and  
the Ising fields by $\phi_m^\ell$ in the notation of section 2.3. 
The index $\ell$ in $\phi_m^\ell$ can be omitted because it is determined by
$m$: indeed, the three Ising fields, the identity, the parafermion 
and the spin are, respectively,
$\phi^0_0=I$, $\phi^0_2=\psi$ and $\phi^1_1=\s$, with periodicities,
$\phi^0_{m+4}=\phi^0_m$ and $\phi^1_{m+2}=\phi^1_m$.
The electron excitation is associated to the parafermion, 
obeying $\psi\times\psi=1$,
and to the $\wh{SU(n)}_1$ field with smallest charge $\b=1$ (with 
Abelian fusion rules over $\b$ modulo $n$) \cite{cvz}.
The other fusion rules with the parafermion field are given by
$\psi\times\psi=I$ and $\s\times\psi=\s$.

The excitations are thus associated with triplets $(\a,\b,m)$, where
the index
$\b$ is mod $n$ and $m$ is mod $4$ $(2)$ if even (odd).  The condition
of integer statistics with the electron $(\a_0,1,2)$ should be
independently computed for the three Ising sectors $m=0,1,2$, leading to
conditions that can be summarized into:
\be
\a\a_0=\frac{2nN+2\b+n\d_{m,1}}{2n}\ , \quad
\a_0^2=\frac{nM+1}{n}\ ,
\qquad N,M\ {\rm integers}.
\label{bs-inte}
\ee
The resulting spectrum is,
\ba
Q&=&\frac{2nN+2\b+n\d_{m,1}}{2nM+2}=\frac{q}{2nM+2}\ ,
\nl
J&=&\frac{\b(n-\b)}{2n}+h_m+\frac{q^2}{4n(2nM+2)}\ .
\label{bs-spec}
\ea
This spectrum identifies the number of $\u1$ sectors and the value
of the charge denominator,
$p=2n\hat{p}$ and $\hat{p}=2nM+2$, respectively. The parity rules
relating the neutral and charge sectors are:
\ba
q&=& 2\b \qquad\qquad {\rm mod}\ 2n, \ m\ {\rm even},
\nl
q&=& 2\b +n \ \qquad {\rm mod}\ 2n, \ m\ {\rm odd}.
\label{bs-parity}
\ea
In particular, the condition on half-integer electron spin
requires that $M$ is odd.

As in the previous cases, the extended characters should resum the spectra
obtained by adding any number of electrons to each anyon:
in the Bonderson-Slingerland states, the addition of one electron
amount to the following index shifts,
\be
(q,\b,m)\ \to\ (q+\hat{p},\b+1,m+2)\ .
\label{bs-shift}
\ee
Therefore, we are led to consider the extended characters,
\ba
\theta_{q,m}&=&\sum_{b=1}^{2n} K_{q+b\hat{p}}\left(\t,2n\z;2n\hat{p} \right)
\ \Theta_{q/2+b}\ \c^0_{m+2b}, \qquad m=0,2,
\nl
\theta_{q,m}&=&\sum_{b=1}^{2n} K_{q+b\hat{p}+n}\left(\t,2n\z;2n\hat{p} \right)
\ \Theta_{q/2+b}\ \c^1_m, \qquad m=1,
\label{bs-char}
\ea
where $q$ is even. In this equation the factors $K_a=K_{a+p}$ and
$\c^\ell_m=\c^\ell_{m+4}$ denote the $\u1$ and $\Z_2$
parafermion characters introduced earlier, respectively, 
and the $\Theta_\b=\Theta_{\b+n}$ are the $\wh{SU(n)}_1$
characters described in \cite{cvz}.
From the periodicity property $\theta_{q+\hat{p},m}=\theta_{q,m+2}$,
the identity of sectors $m=1\sim 3$, and the $V$ transformation of
$K$ characters, we obtain the following 
values for the topological order and filling fraction of the
Bonderson-Slingerland states:
\be
T.O. =3(nM+1), \qquad \nu=2+ \frac{n}{nM+1},\qquad M\ {\rm odd}, 
\label{bs-to}
\ee
that reduce to those of the Pfaffian state in the $n=1$ case.

Owing to the fact that $(n,nM+1)=1$, a representative set of
the $(nM+1)$ values of the charge $q$ in the extended characters 
can be chosen to be $q=2an$, $a=0,1,\dots,nM$, leading to the characters:
\ba
\theta_{a,0}&=&
\sum_{b=1}^{2n} K_{2an+b\hat{p}}\left(\t,2n\z;2n\hat{p} \right)
\ \Theta_b \left( I\ \d^{(2)}_{b,0} + \psi\ \d^{(2)}_{b,1} \right) , \qquad m=0,
\nl
\theta_{a,1}&=&
\sum_{b=1}^{2n} K_{(2a+1)n+b\hat{p}}\left(\t,2n\z;2n\hat{p} \right)
\ \Theta_b\ \s\ , \qquad\qquad\qquad m=1,
\nl
\theta_{a,2}&=&
\sum_{b=1}^{2n} K_{2an+b\hat{p}}\left(\t,2n\z;2n\hat{p} \right)
\ \Theta_b \left(\psi\ \d^{(2)}_{b,0} + I\ \d^{(2)}_{b,1} \right) , \qquad m=2,
\nl
&&\qquad a=0,1,\dots,nM.
\label{bs-char2}
\ea
where we rewrote the Ising characters with the same symbols of the fields.

The annulus partition function is therefore given by the 
diagonal combination of these characters,
\be
Z_{\rm annulus}^{\rm BS} =
\sum_{a=0}^{nM}
\ \left\vert\ \theta_{a,0}\ \right\vert^2\ +
\ \left\vert\ \theta_{a,1}\ \right\vert^2\ +
\ \left\vert\ \theta_{a,2}\ \right\vert^2\ .
\label{bs-zann}
\ee
In particular, the expression earlier found for the Pfaffian
state (\ref{ising-zann}) is recovered for $n=M=1$.

As before, the final step is to verify that this set of extended characters
is closed under $S$ modular transformation, that is unitarily represented.
As discussed in Appendix A in more detail, the transformation of Ising
characters is particularly simple in the following basis,
\ba
&&\wt{\c}_m=\left\{\frac{I+\psi}{\sqrt{2}},\s , \frac{I-\psi}{\sqrt{2}} 
 \right\},
\nl
&&\frac{1}{\sqrt{2}}\left(I\left(-\t^{-1}\right)
+\psi\left(-\t^{-1}\right)
\right)=
\frac{1}{\sqrt{2}}\left(I(\t)+\psi(\t) \right),
\nl
&&\s\left(-\t^{-1}\right) = \frac{1}{\sqrt{2}}\left(I(\t)-\psi(\t) \right).
\label{bs-s1}
\ea
The full $S$ matrix in this basis turns out to be:
\be
S_{(a,m),(a',m')}= \frac{e^{i2\pi \frac{aa'n}{nM+1}}}{\sqrt{nM+1}} 
\left(
\begin{array}{lll}
1 & 0 & 0 \\
0 & 0 & e^{i2\pi a'n/2(nM+1)} \\
0 & e^{i2\pi an/2(nM+1)} & 0
\end{array}
\right),
\label{bs-s2}
\ee
that is unitary and again factorized in neutral and charged parts, up to
details.

\subsection{Non-Abelian spin-singlet states}

The main physical idea of this proposal is that of generalizing
the clustering property of the Read-Rezayi states to spinful electrons:
namely, of requiring that the ground-state wave function does not vanish when
$k$ electrons with spin up and $k$ with spin down are brought to
the same point \cite{nass}.
In terms of CFT operator product expansion, we need two species of
parafermions, $\psi_\uparrow,\psi_\downarrow$, that obey 
$(\psi_\uparrow)^k=(\psi_\downarrow)^k=I$. 
This possibility is offered by the generalized parafermions that are
obtained by the coset construction $\wh{SU(3)}_k/\wh{U(1)^2}$,
first discussed in \cite{gepner}. Let us recall their main features.

The $\wh{SU(3)}_k/\wh{U(1)^2}$ parafermion fields $\phi^\L_\l$
are characterized by the pair of $SU(3)$ weights $(\L,\l)$, 
that belong to the two-dimensional lattice generated by
the positive fundamental weights $\mu_1,\mu_2$, with scalar products 
$(\mu_1,\mu_1)=(\mu_2,\mu_2)=2/3$ and $(\mu_1,\mu_2)=1/3$.
The dual lattice is generated by the positive roots $\a_1,\a_2$,
with $(\a_1,\a_1)=(\a_2,\a_2)=2$ and $(\a_1,\a_2)=-1$,
i.e. $\a_1=2\mu_1-\mu_2$ and $\a_2=-\mu_1+2\mu_2$. 
The weight $\L$ takes values inside the so-called truncated Weyl chamber,
$\L\ \in P^+_k$ , while $\l$ is a vector of the
weight lattice $P$ quotiented by the $k$-expanded root lattice
$Q$ , $\l\in P/kQ$. In more detail, we have the following values of 
the weights $(\L,\l)$ and integer labels $(n_1,n_2,\ell_1,\ell_2)$:
\be
\begin{array}{ll}
\phi^\L_\l \equiv\phi^{n_1,n_2}_{\ell_1,\ell_2},&
\\
 \L=n_1\mu_1+n_2\mu_2, & 0\le n_1,n_2,\  n_1+n_2 \le k,
\\
 \l=\ell_1\mu_1+\ell_2\mu_2, & (\ell_1,\ell_2) \ {\rm mod}\ (2k,-k),(k,-2k),
\\
 & n_1-n_2=\ell_1-\ell_2 \ \ {\rm mod}\ 3.
\end{array}
\label{su3-w}
\ee
The last condition in this equation states that the $SU(3)$ triality 
of two weights is the same, $(\L-\l)\in Q$.
Another trivalent condition is the so-called coset field identification 
\cite{gepner}, saying that the following labellings are equivalent:
\be
\phi^{n_1,n_2}_{\ell_1,\ell_2} = 
\phi^{k-n_1-n_2,n_1}_{\ell_1+k,\ell_2} =
\phi^{n_2,k-n_1-n_2}_{\ell_1,\ell_2+k}\ .
\label{f-id}
\ee
The number $k^2(k+1)(k+2)/6$ of parafermion fields is 
easily determined from these data: the product of 
the independent $\L$ values is $(k+1)(k+2)/2$, that of the $\l$'s ones is
$3k^2$ (from the areas of lattice fundamental cells 
$|k\a_1 \wedge k\a_2 |/|\mu_1\wedge \mu_2|$), while
the two conditions account for a factor $1/9$.

Within this theory, the two fundamental parafermions are
$\psi_\uparrow=\phi^0_{\a_1}$ and $\psi_\downarrow=\phi^0_{-\a_2}$,
which have Abelian fusion rules with all fields: actually, 
the lower index $\l$ of $\phi^\L_\l$ belongs to an Abelian
charge lattice and is thus additive modulo $kQ$, leading to the
fusion rules
$\phi^\L_\l \times \phi^0_{\a_i}=\phi^\L_{\l+\a_i\ {\rm mod} \ kQ}$.
(The choices of root sign in the definition of the fundamental parafermions
will be relevant in (\ref{nass-2e}).)

The conformal dimension of parafermion fields are given by
\cite{qiu},
\be
h^\L_\l=\frac{(\L,\L+2\r)}{2(k+3)}- \frac{|\l|^2}{2k}\ ,
\label{su3-h}
\ee
where $\rho=\mu_1+\mu_2$ is half the sum of positive roots and
$|\l|^2=(\l,\l)$;
finally, the central charge of the theory is $c=8k/(k+3)$.

The excitations of the NASS state are characterized by
two Abelian quantum numbers, the charge $Q$ and the intrinsic
spin $S$ (not to be confused with the orbital $J$, equal to 
half the statistics). The full description is based on the RCFT 
$\left(\wh{SU(3)}_k/\wh{U(1)^2}\right)\otimes \u1_n\otimes\u1_p$:
in the following, we shall determine the spin and charge
compactification parameters $(n,p)$ and the selection
rules relating the $(S,Q)$ values to the non-Abelian weights $(\L,\l)$
by extending the conditions (A)-(D) for physical excitations
of section 2.2.
The NASS quasiparticles are described by parafermion
fields and vertex operators with $\a$ and $\b$ charges for electric charge and
spin, respectively:
\be
\Psi_i=\phi^\L_\l\ e^{i\a\vf}\ e^{i\b\vf'}\ ;
\label{nass-ex}
\ee
in particular, the two electron fields are,
\be
\Psi_e^\uparrow=\phi^0_{\a_1}\ e^{i\a_0\vf}\ e^{i\b_0\vf'}\ ,
\qquad
\Psi_e^\downarrow=\phi^0_{-\a_2}\ e^{i\a_0\vf}\ e^{-i\b_0\vf'}\ .
\label{nass-e}
\ee
 
The conditions of integer statistics of each excitation with the
two electrons read:
\ba
&&\a\a_0+\b\b_0 +h^\L_{\l+\a_1}-h^\L_\l-h^0_{\a_1} \in \Z \ ,
\nl
&&\a\a_0-\b\b_0 +h^\L_{\l-\a_2}-h^\L_\l-h^0_{-\a_2} \in Z \ ;
\label{nass-stat}
\ea
from the conformal dimensions (\ref{su3-h}),
one finds that the weight $\L$ does not enter in these conditions.
We now analyze the various cases of mutual statistics in turn:
for the electrons with/among themselves, we obtain,
\be
\a_0^2-\b_0^2=\frac{1}{k}+N, \qquad
\a_0^2+\b_0^2=\frac{2}{k}+N';
\label{nass-2e}
\ee 
for the electrons with a quasiparticle with labels 
$(\ell_1,\ell_2,\a,\b)$, they are,
\be
2\a\a_0=M+ M'+\frac{\ell_1-\ell_2}{k}, \qquad
2\b\b_0=M-M'+\frac{\ell_1+\ell_2}{k}\ ,
\label{nass-ei}
\ee 
with $N,N',M,M'$ integers.

Applying these quantization conditions, we obtain the spectrum:
\ba
Q&=& \frac{\a}{\a_o}=\frac{(M+M')k +\ell_1-\ell_2}{3+2kN} =\frac{q}{3+2kN},
\qquad N \ {\rm odd},
\nl
S&=& \frac{\b}{2\b_o}=\frac{(M-M')k +\ell_1+\ell_2}{2} =\frac{s}{2},
\nl
J&=& h^\L_\l +\frac{\a^2+\b^2}{2}=h^\L_\l+\frac{q^2}{4k(3+2kN)}
+\frac{s^2}{4k}\ .
\label{nass-spec}
\ea
In these equations, the condition of electron intrinsic spin $S=\pm 1/2$ 
has fixed $N=N'$ (i.e. $\b_0^2=1/2k$), and that of half-integer orbital spin 
$J$ has selected $N$ odd.

From the spectrum, we identify the compactification parameters for $s$ and $q$,
\be
s \quad {\rm mod}\ n=2k,\qquad\qquad q\quad {\rm mod}\ p=2k\hat{p}, \ 
\hat{p}=3+2kN,
\label{nass-per}
\ee
and the parity rules,
\be
\ell_1=\frac{s+q}{2}\quad {\rm mod}\ k, \qquad
\ell_1=\frac{s-q}{2}\quad {\rm mod}\ k,
\label{nass-pr}
\ee
that also imply $s=q$ mod $2$.
In summary, a quasiparticle is characterized by the integer labels 
$(\L,\l,q,s)\equiv (n_1,n_2;\ell_1,\ell_2;q,s)$ obeying these parity rules and 
further constrained by triality, $n_1-n_2=\ell_1-\ell_2=q$ mod $3$.

The addition of electrons to a quasiparticles causes the following
index shifts:
\be
\begin{array}{l|llll}
& \D \ell_1& \D \ell_1& \D q &\D s \\
\hline
+\psi^\uparrow & 2 & -1& \hat{p} & 1
\\ 
+\psi^\downarrow & 1 & -2& \hat{p} & -1
\end{array}
\label{nass-shift}
\ee
The extended characters are thus obtained by products of
Abelian characters, $K^{(Q)}=K(\t,2k\z;2k\hat{p})$, 
$K^{(S)}=K(\t,0;2k)$, for charge and intrinsic spin, respectively,
and of parafermion characters $\c^\L_\l(\t)$, summed over
electron excitations:
\be
\theta^\L_{q,s}=\sum_{a,b} \ K^{(Q)}_{q+(a+b)\hat{p}}\ 
K^{(S)}_{s+a-b}\ \c^{n_1,n_2}_{\ell_1,\ell_2}\ ,
\label{nass-th1}
\ee
where the values of $(\ell_1,\ell_2)$ are constrained by the 
parity rules (\ref{nass-pr}).

Further specifications/conditions on the expression (\ref{nass-th1})
are the following:

i) the $(a,b)$ ranges are fixed by checking the periodicity of the
summand; upon inspection, the ranges $a,b$ mod $2k$ are surely periodic,
but also shifts by $(a,b)\to(a+k,b+k)$ and $(a,b)\to(a+k,b-k)$ maps
the sums into themselves: thus, we can take $a=1,\dots,k$ and
$b=1,\dots, 2k$.

ii)
The parity rule (\ref{nass-pr}) has $3$ solutions for $\l\in P/kQ$, i.e. for
$(\ell_1,\ell_2)$ mod $(2k,-k),(k,-2k)$; these are,
\be 
\left({ \ell_1 \atop \ell_2} \right)
=\left( {\frac{s+q}{2} \atop  \frac{s+q}{2} }\right), \ \
\left( {\frac{s+q}{2} +k \atop  \frac{s+q}{2} }\right), \ \
\left( {\frac{s+q}{2} \atop  \frac{s+q}{2} +k }\right).
\label{l-sol}
\ee
However, using the field identifications (\ref{f-id}), 
these three solutions can be traded for $\L$ changes and should 
not be considered as independent.
We thus obtain:
\be
\theta^\L_{q,s}=\frac{1}{2}\sum_{a,b=1}^{2k} \ K^{(Q)}_{q+(a+b)\hat{p}}\ 
K^{(S)}_{s+a-b}\ \c^\L_{\frac{s+q}{2} +2a+b,\frac{s-q}{2} -a -2b }\ .
\label{nass-th2}
\ee

iii)
Independent values of $(q,s)$ indeces, 
$q=s$ mod $2$, are found by checking the periodicities of $\theta^\L_{q,s}$;
we find that:
\be
\theta^\L_{q,s+2}= \theta^\L_{q,s}\ , \qquad
\theta^{n_1,n_2}_{q+\hat{p},s+1}= \theta^{k-n_1-n_2,n_1}_{q,s}\ .
\label{ch-per}
\ee
Therefore, the intrinsic spin index is not independent and
can be taken to be $s=0$ $(1)$ for
$q$ even (odd), while the charge range is $q=1,\dots,\hat{p}=3+2kN$.

The number of independent extended NASS characters is
given by the values of $q$ and $\L$ constrained by 
triality, $n_1-n_2=q$ mod $3$. We obtain,
\be
T.O.= (3+2kN)\frac{(k+1)(k+2)}{6},\qquad \nu=\frac{2k}{2kN+3}\ ,\quad
N \ {\rm odd},
\label{nass-to}
\ee  
where the filling fraction follows from the $V$ transformation of $K^{(Q)}$.

In conclusion, the NASS partition function on the annulus geometry is,
\be
Z_{\rm annulus}^{\rm NASS} =
\sum_{q=1}^{2kN+3}\sum_{s=0,1 \atop s=q\ {\rm mod}\  2}\  
\sum_{ 0\le n_1+n_2\le k \atop n_1-n_2=q\ {\rm mod} \ 3}
\ \left\vert\ \theta_{q,s}^{n_1,n_2}\ \right\vert^2\ .
\label{nass-zann}
\ee

The modular transformation of the NASS characters is derived
in Appendix A and reads:
\be
\theta^\L_{q,s}= \d^{(2)}_{q,s} \ \d^{(3)}_{n_1-n_2,q}
\sum_{q'=1}^{\hat{p}}\sum_{s'=0,1 \atop s'=q'\ {\rm mod}\  2}\
\sum_{ \L'\in P^+_k} \frac{1}{\sqrt{\hat{p}}}\ 
e^{-i2\pi \frac{qq'N}{2\hat{p}}}\ \d^{(2)}_{q',s'}\ s_{\L,\L'}\ 
\theta^{\L'}_{q',s'}\ ,
\label{nass-s}
\ee
where $s_{\L,\L'}$ is the $\wh{SU(3)}_k$ modular $S$-matrix.

\subsubsection{\bf The $k=2$, $M=1$ case.}

Let us discuss the simplest NASS state with $k=2$ and $M=1$, corresponding
to $\nu=4/7$. There are $8$ parafermionic fields,
\ba
&&  I =\Phi^{0,0}_{0,0} ,\ \ \
  \psi_{1}=\phi^{0,0}_{2,-1},\ \ \
  \psi_{2}=\phi^{0,0}_{-1,2}=\phi^{0,0}_{1,-2}  ,\ \ \
  \psi_{12}=\Phi^{0,0}_{1,1}=\Phi^{0,0}_{3,-3}  ,
\nl
&&
  \sigma_{\da}=\Phi^{0,1}_{0,1}= \Phi^{1,1}_{2,-1}  , \ \ \
  \sigma_\ua =\Phi^{1,0}_{1,0} = \Phi^{1,1}_{-1,2} ,\ \ \
  \sigma_3 =\Phi^{1,1}_{1,1}  ,\ \ \
  \rho=\Phi^{1,1}_{0,0}  ,
\label{nass-p}
\ea
that are also written in the notation of Ref. \cite{ardonne}. 

There are $14$ sectors in the theory, and corresponding extended characters
$\theta^\L_\l$, that are made of the parafermion characters $\c^\L_\l$
combined with Abelian characters $K^{(Q)}_m K^{(S)}_s$, with $m$ mod $p=28$
and $s$ mod $4$; quasiparticles have charge $m/7+4\Z$ and intrinsic spin 
$s/2+2\Z$.  
The partition function (\ref{nass-zann}) can be rewritten:
\ba
&&Z^{\rm NASS}_{annulus}(k=2) \ =
\nl
&&=
\sum_{m=0,4,8,12,16,20,24}\left\vert
\chi_I \KK_{m,0}+ \chi_{\psi_1}\KK_{m+7,1}+
\chi_{\psi_2}\KK_{m+7,3} + \chi_{\psi_{12}} \KK_{m+14,0}
\right\vert^2
\nl
&&+
\sum_{m=0,4,8,16,20,24,26}\left\vert
\chi_{\rho} \KK_{m,0}+ \chi_{\s\downarrow} \KK_{m+7,1}+
\chi_{\s_\uparrow}\KK_{m+7,3} +\chi_{\s_3} \KK_{m+14,0}
\right\vert^2,
\label{z-nass2}
\ea
where we denoted,
\begin{equation}
\KK_{m,s} = \KK_{m+14,s+2}= K^{(Q)}_mK^{(S)}_s + K^{(Q)}_{m+14}K^{(Q)}_{s+2}\ .
\label{nass-kk}
\end{equation}
The $m=0$  term in the first sum contains the identity $I$ and the two 
electron excitations made of $\psi_1$ and $\psi_2$ parafermions,
obeying the fusions $(\psi_1)^2=(\psi_2)^2=I$ and also
$\psi_1\times\psi_2=\psi_{12}$, leading to the forth term in that sector.
The term $m=8$ in the second sum contains the basic quasiparticle
made of spin fields $\s_{\uparrow},\s_{\downarrow}$, with smallest charge 
$Q=1/8$ and spin $S=\pm 1/2$.
The other terms are further quasiparticle excitations. 
Note that the division into extended characters is in agreement with
the fusion subalgebras made by multiple fusing of $\psi_1$, $\psi_2$ 
with all the fields; the complete
table of fusion rules can be found in \cite{ardonne}.

\section{Physical applications of partition functions}

Besides providing a complete definition of the Hilbert space of
edge excitations, their quasiparticle sectors and fusion rules,
the partition functions are useful for 
computing physical quantities that could be measured in
current experiments on non-Abelian
statistics: in particular Coulomb blockade \cite{cb0} and entropy  
measurements through the thermopower effect \cite{thermop}.
We first recapitulate the main features of partition functions
discussed in the previous sections, using a standard
notation for all of them (inspired by the Read-Rezayi states).
The partition function on the annulus is a sum of squared terms:
\be
Z_{\rm annulus}=\sum_{a,\ell} \left\vert \theta^\ell_a\right\vert^2\ ,
\label{z-st}
\ee
where $a$ and $\ell$ are the Abelian and non-Abelian indices, 
respectively, possibly obeying some relative conditions.
The total number of terms in
the sum is equal to the topological order of the Hall state.
Each extended character takes the form,
\be
\theta^\ell_a= K_a \c_{m}^\ell +K_{a+\hat p} \c^\ell_{m+\hat m}+\cdots\ ,
\label{theta-st}
\ee 
where the Abelian indeces $(a,m)$ are related by a selection rule
(parity rule) and the various terms in the sum describe the addition 
of electrons (quantum numbers $(\hat{p},\hat{m})$) 
to the basic quasiparticle of that sector.

The extended characters transforms linearly under the modular
$S$ transformation, 
\be
\theta_a^\ell\left(-1/\t\right) =\sum_{a',\ell'}\ S_{aa'}\ s_{\ell\ell'}\ 
\theta_{a'}^{\ell'}(\t)\ ,
\label{s-st}
\ee
where the $S$-matrix is basically factorized into an Abelian
phase, $S_{aa'}\sim\exp\left(i2\pi aa' N/M\right)$, and a less trivial
non-Abelian part $s_{\ell\ell'}$.

Finally, the expression of charged characters 
$K_a(\t,n\z;n\hat{p})$ is well-known and given by the theta functions 
(\ref{thetaf}) 
with parameters defined in (\ref{t-z-def}), while that of non-Abelian ones 
$\c^\ell_m(\t_n)$ is less explicit: nonetheless, the knowledge of their
leading low-temperature behavior $\t_n\to i\infty$
is usually sufficient:
\be
\c^\ell_m (\t_n) \sim d^\ell_m \ 
e^{i2\pi \t_n \left(h^\ell_m -c/24 \right)}\ ,
\qquad \I \t_n=\frac{\b}{2\pi} \frac{v_n}{R}\ . 
\label{chi-exp}
\ee
In this equation, $h^\ell_m$ is the conformal dimension of the corresponding
non-Abelian field and $c$ is the central charge of the CFT; $d^\ell_m$ is
the multiplicity of the low-lying state in that sector,
that is present in some theories due to their extended symmetry.
We also changed the modular parameter $\t\to\t_n$ to account for a 
different Fermi velocity $v_n$ of neutral excitations.

\subsection{Coulomb blockade in quantum Hall droplets}

The study of Coulomb blockade current peaks has been initiated
in \cite{cb0}, where they were shown to
provide interesting information on the spectrum of edge excitations. 
Indications of experimental feasibility have 
been reported in \cite{cb-exp}.
The physical system is an isolated droplet of Hall fluid and the
current peaks are due to
electrons that tunnel in and out the system, one by one because
the voltage bias $V_o$ is counterbalanced by the electrostatic 
energy $E_C$ of the droplet; the latter is relevant for
small droplets of size $2-20$ $\mu m$.
We shall discuss the cases of $T=0$ and $T>0$, both at and out of 
equilibrium.

\subsubsection{\bf $T=0$.}

At very small temperatures and low bias $V_o\sim 0$, the conductance peaks 
can only occur for exact energy matching: 
the spectrum of droplet edge excitations should possess degenerate
energy levels under addition of one electron, $Q\to Q+1$.
The use of the partition function at $T=0$ has already been
discussed in earlier publications \cite{cgz}\cite{cvz} 
and will be briefly summarized here.

The discrete spectrum, such e.g. (\ref{spec-h}) in the Read-Rezayi state,
can be continuously deformed by varying the size of the droplet,
which modifies the background charge and adds a capacitive energy to
the $\u1$ part of the spectrum, $E\propto (Q-Q_{\rm bkg})^2$.
In the character $K_\l(\t,\z;p)$, the deformed energy is:
\be
E_{\l,\s}(n) = \frac{v}{R} \frac{\left(\l+p n-\s \right)^2}{2p} \ , 
\qquad \s= \frac{ B \D  S}{\phi_o} \ ,
\label{def-en} 
\ee
where $\s$ is a dimensionless measure of area deformation.
The neutral part of the spectrum is unaffected.
In the sector $\theta^\ell_a$ (\ref{theta-st}), corresponding to the partition 
function on disk geometry with the $(\ell,a)$ quasiparticle in the bulk,
one should compare the lowest energies
on pairs of consecutive terms, i.e. with consecutive electron numbers,
and obtain the values $\s=\s^\ell_i$ for energy matching at which
the current can occur. 
One finds that the distance among them $\D \s^\ell_i$ 
is not constant owing to the contribution of neutral energies,
\be
\D \s_i^\ell=\s^\ell_{i+1}-\s^\ell_i=\frac{1}{\nu}+ 
\frac{v_n}{v}\left(
h^\ell_{a+2i+2}-2 h^\ell_{a+2i}+h^\ell_{a+2i-2} \right)\ .
\label{pf-diff}
\ee
For example, in the $\Z_k$ Read-Rezayi states, 
the following peak patterns were obtained \cite{cvz},
\ba
\ell &=& 0,k : 
\nl
&&
\ \ \D\s=\left(\D+2r,\D,\cdots,\D\right), 
\ \ \ \ \ \ \ \ \ (k)  \ \ {\rm groups},
\nl
\ell &=&1,\dots,k-1 :
\nl
&&
\ \ \D\s=\left(\D+r,\D,\cdots,\D+r,\D,\dots,
\D\right), \ \ \ \ (\ell) (k-\ell)\ \ {\rm groups},
\nl
&&\ \ 
\D = \frac{1}{\nu}-\frac{v_n}{v}\frac{2}{k}\ ,
\qquad  r=\frac{v_n}{v}\ ,
\label{pf-patt}
\ea
they depend on the non-Abelian index $\ell$ (counting the number of
basic quasiparticle $\s_1$ in the bulk).
The modulation of peak distances is equal to the ratio 
of velocities of neutral to charged excitations $v_n/v=r\sim 1/10$.
Coulomb blockade peaks are also obtained by tuning the magnetic field
but their pattern 
is the same once expressed in terms of flux change \cite{cvz}.
Let us stress that this is a equilibrium phenomenon caused by
adiabatic variation of the Hamiltonian.
The analysis of peak patterns for the non-Abelian theories
discussed in the previous sections has been carried out in
\cite{cb0}\cite{dopp}\cite{cvz} and will not be repeated here.

In Ref.\cite{dopp}, it has been argued that the modulation of
Coulomb peaks at $T=0$ is not a signature of non-Abelian states, since
similar or equal patterns can also be found in Abelian states 
that have non-trivial neutral excitations.
Indeed, at $T=0$ one is probing only the leading behavior (\ref{chi-exp}) 
of the neutral characters $\c^\ell_m$, that can be
the same for different theories, Abelian and non-Abelian.
For example, the Read-Rezayi states and the $(331)$ Haldane-Halperin
Abelian hierarchical fluids have this property.

One physical explanation of this fact was given in Ref. \cite{cgt2},
where it was shown that the $(331)$ state can be considered as a ``parent''
Abelian theory of the Read-Rezayi state, in the following sense. 
The Haldane-Halperin theory possesses the same $k$-th electron clustering 
property but it is realized through the addition of a
$k$-fold quantum number, called e.g. ``color''.
Since electrons of different color are distinguishable, their wavefunction
needs not to vanish when $k$ of them come close; this behavior 
can be achieved in a standard Abelian multicomponent 
theory specified by a charge lattice.
In this setting, the Read-Rezayi state is reobtained when electrons
are made indistinguishable, i.e. when wavefunctions are antisymmetrized
with respect to all electrons independently of the color.

In CFT language, such antisymmetrization amounts to a projection
in the Hilbert space that does not change the sectors and selection
rules over which the partition function is built; 
the neutral CFT undergoes a coset reduction, from the Abelian lattice with
$\wh{SU(k)}_1\otimes \wh{SU(k)}_1 $ symmetry to the non-Abelian theory
 $\wh{SU(k)}_1\otimes \wh{SU(k)}_1/ \wh{SU(k)}_2$, the latter being
another realization of $\Z_k$ parafermions \cite{cgt2}. 
Therefore, the Abelian and non-Abelian fluids have the same form of the
partition function, only the neutral characters are
different: however, their leading behavior (\ref{chi-exp}) is unaffected by
the projection, leading to the same patterns of Coulomb peaks at $T=0$.
In conclusion, the presence of equal Coulomb peaks patterns in two
different theories, Abelian and non-Abelian, is not accidental and it may
suggest a physical mechanism behind it.


\subsubsection{\bf $T>0$ at equilibrium.}

The analysis of temperature dependent Coulomb blockade in Hall droplets 
has been initiated in Ref. \cite{cbT} \cite{georgiev}.
From the disk partition function, one obtains
the thermal average of the charge, 
\be
\langle Q\rangle =- \frac{1}{\b} \frac{\de}{\de V_o} \log \theta^\ell_a\ ,
\label{q-exp}
\ee
whose qualitative expression at low temperature is:
\be
\langle Q\rangle_\ell =\frac{\sum_i \ i\ \d_\b\left(\s-\s^\ell_i\right)}
{\sum_i\d_\b\left(\s-\s^\ell_i\right)}\ ,
\label{q-t}
\ee
where the $\d_\b(x)$ is a Gaussian representation of the delta function 
with spread proportional to $1/\b$.
Upon varying the droplet area $\s$  by $\D\s^\ell_i$,
the dominant term in the sum changes from the $i$-th to the $(i+1)$-th one
and the value of $\langle Q\rangle$ jumps by one unit.
Therefore, this quantity has a characteristic staircase shape for electrons
entering the droplet; the effect of the temperature is that of rounding
the corners, that become bell-shaped peaks in the derivative of 
$\langle Q\rangle$ with respect to the control parameter. 

We should distinguish two temperature scales:
\be
T_n=O\left(\frac{v_n}{R}\right)\sim 50\ mK \ ,\qquad
T_{ch}=O\left(\frac{v}{R}\right)\sim 250\ mK \ ,
\label{t-def}
\ee
that correspond to typical energies of neutral and charged excitations
of a small droplet ($R\sim 10\ \mu m$), respectively.
The discussion above applies to the range $T<T_n$, where one sees
the smoothening of the $T=0$ peaks.

A new feature \cite{cgz}\cite{cvz}\cite{georgiev} is associated to
the multiplicity factor $d^\ell_m$ in the leading behavior of the
neutral characters (\ref{chi-exp}): if $d^\ell_m>1 $, the electron entering
the droplet finds more than one available degenerate state.
At equilibrium for $T \simeq 0$, the probability of one-electron 
tunneling under parametric variation of the Hamiltonian is either one 
(at degeneracy) or zero (off degeneracy),
thus the presence of more than one available empty state
is not relevant.
Inclusion of a small finite-size energy splitting among the $d^\ell_m$ states
does not substantially modify the peak shape.
In particular, the earlier analysis of Refs. \cite{cgz}\cite{cvz},
indicating the possibility of a comb-like sub-structure of peaks for 
$d^\ell_m>1$ is not actually correct at equilibrium (there is an off-equilibrium
effects to be discussed in next section).
In conclusion, the presence of level multiplicities does not influence
the $T=0$ peak pattern. 

On the other hand, for $T>0$ these multiplicities 
lead to a displacement of the peak centers, as follows\cite{georgiev}: 
\be
\s_i^\ell\to \s_i^\ell +\frac{T}{T_{ch}}
\log \left( \frac{d^\ell_{a+2i}}{d^\ell_{a+2i+2}} \right) \ ,
\label{mult-g}
\ee
as is clear by exponentiating the $d^\ell_m$ factor into the energy.
The effect is observable for $T\le T_n$
by increasing the experimental precision.
In particular, the multiplicities and peak displacements are present
in the $(331)$ states (due to color multiplicity) and absent in the
Read-Rezayi states, thus providing a difference in the Coulomb peaks
of these two theories for $T>0$ \cite{georgiev}.

The level multiplicities of the other non-Abelian states
are found by the leading expansion of the
respective neutral characters:
\be
\begin{array}{lll}
d^\ell = \left(\ell+1\right)\ ,& \c^\ell\sim d^\ell\ q^{h^\ell-c/24} \ ,
& SU(2)\ {\rm NAF \ and}\ \ov{\rm RR}\ ,
\\
d_\b = \left({n \atop \b}\right) \ , &\Theta_\b \sim d_\b\ 
q^{\b(n-\b)/2n -n/24}\ ,
& SU(n)\ {\rm BS \ and\ Jain}\ ,
\nl
d^\ell_m= d^\L_\l = 1\ , && {\rm RR\ and \ NASS}.
\end{array}
\label{na-mult}
\ee

The peak displacements are particularly interesting when they can be used to
distinguish between competing theories with same filling fraction
and Coulomb peak pattern at $T=0$.
In the case of the Jain hierarchical states, one can test
the multiplicities of states predicted by the multicomponent 
$\wh{SU(n)}_1\otimes\u1$ Abelian theories versus
the $\winf$ minimal models possessing no degeneracy, as discussed
in earlier works \cite{w-min}\cite{cz2}.
The peak pattern including displacements of the $\wh{SU(n)}_1\otimes\u1$
theory are, for $\nu=n/(2sn\pm 1)$:
\ba
\D\s_i \! &=& \!  \frac{1}{\nu}- \frac{v_n}{v}\frac{1}{n}
+\frac{T}{T_{ch}}\log\frac{(i+1)(n-i+1)}{i(n-i)}\ ,
\qquad  i=1,\dots,n-1,
\cr
\D\s_n \!  &=& \!  \frac{1}{\nu}+ \frac{v_n}{v}\frac{n-1}{n}
+\frac{T}{T_{ch}}\log\frac{1}{n^2}\ .
\label{dis-peak} 
\ea 
The $\winf$ minimal models possess the same pattern without the
temperature dependent term for $T<T_n$.
The formula (\ref{dis-peak}) supersedes our earlier analyses of 
this problem \cite{cgz}\cite{cvz}.

Another interesting case is the competition between the Pfaffian state
at $\nu=5/2$ and its particle hole conjugate state $\ov{\rm RR}$, possessing
multiplicities due to its $SU(2)$ symmetry, as indicated in (\ref{na-mult}).
Let us discuss this point in some detail. 
The Coulomb peak distances in the $\ov{\rm RR}$
states, parameterized by $(\ell,a)$, $a=\ell$ mod $2$
(cf. (\ref{arr-theta})), have period two for any $k$ and their expressions
including the  shift due to level multiplicity are, for
$\nu=2/(2M+k)$:
\ba
\D\s^\ell_a &= &\frac{1}{\nu} + (-1)^a \left[
\frac{v_n}{v}\frac{2\ell -k}{2}
+ 2\frac{T}{T_{ch}} \log\left(\frac{k-\ell+1}{\ell+1}\right) \right] ,
\nl
&&\qquad \ell=0,1,\dots,k .
\label{arr-cb}
\ea

In particular, for $\nu=5/2$ ($k=2$) there is the so-called
even/odd effect \cite{cb0} for both the RR, the $\ov{\rm RR}$ 
and $(331)$ states
(no modulation for $\ell=1$, pairwise modulation for $\ell=0,2$).
The peak shifts are present in the last two theories for $\ell=0$,
but are absent in the Pfaffian state.
It follows that the observation of a temperature dependent
displacement in the position of paired peaks could support both 
the $(331)$ and the anti-Pfaffian  at $\nu=5/2$
(the last case was not discussed before).

At higher temperatures, in the region $T_n<T<T_{ch}$, 
a new feature appears:
the Boltzmann factors relative to higher neutral excitations 
can be of order one and can contribute beyond the leading
$T\to 0$ term in (\ref{chi-exp}).
As observed in \cite{cbT}, it is convenient to perform the $S$ modular
transformation on the neutral characters $\c(\t_n)$, 
$\t_n\to -1/\t_n\sim iT/T_n$, and expand them for $\t_n\to 0$.

Using again the Read-Rezayi case as an example,
we find from (\ref{pf-th}),(\ref{pf-th-s}) upon keeping the first three terms,
\ba
\!\!\!\!\!\!\!\!\!\! \!\!\!\!\!
\c^\ell_m (\t_n) &\sim &\frac{1}{\sqrt{2k}}\left(
s_{\ell 0}\ \c^0_0\left(\frac{-1}{\t_n}\right) 
+ s_{\ell 1}\ e^{-i\pi m/k}\ \c^1_1 \left(\frac{-1}{\t_n}\right)  +
s_{\ell 1}\ e^{i\pi m/k}\ \c^1_{-1} \left(\frac{-1}{\t_n}\right) 
\right) 
\nl
&\sim& s_{\ell 0} \left(
1 + \frac{s_{\ell 1}}{s_{\ell 0}}
e^{-i(2\pi/\t_n)h^1_1 }\ 2 \cos\left(\frac{\pi m}{k}\right) 
\right)
\nl 
&\sim & s_{\ell 0}\ e^{D^\ell_m}\ , \qquad \qquad\t_n\to 0.
\label{s-neut}
\ea
In this limit, the extended characters (\ref{pf-th}) become:
\be
\theta^\ell_a \sim
\sum_{b=1}^k K_{a+b\hat p}(\t)\ s_{\ell 0}\ e^{D^\ell_{a+2b}}\ ,
\label{ps-corr}
\ee
where,
\be
D^\ell_{a+2b} = e^{-4\pi^2 h^1_1 T/T_n}\ 2\cos\frac{\pi(\ell+1)}{k+2}
\ 2\cos\frac{\pi(a+2b)}{k} \ .
\label{rr-d}
\ee
The distance between  Coulomb peaks in RR states 
for $T>T_n$ is therefore given by (cf. (\ref{pf-diff})):
\be
\D\s^\ell_i=\frac{1}{\nu} -\frac{T}{T_{ch}}
e^{-4\pi^2 h^1_1 T/T_n}\ 8 \cos\frac{\pi(\ell+1)}{k+2}
\cos\frac{\pi(a+2i)}{k}\left(\cos\frac{2\pi}{k}-1  \right) .
\label{pf-tn}
\ee
Although exponentially small, this temperature effect is full-fledged
non-Abelian, since it involves the ratio $s_{\ell 1}/s_{\ell 0}$ of components
of the $S$-matrix for neutral states; such ratios also characterizes other
non-Abelian probes, including the most popular Fabry-Perot interference phase
\cite{na-interf}.

The corresponding correction terms for the other Hall states  are,
\ba
SU(n)\ \mathrm{Jain}: & &
D_b = e^{-4\pi^2 h_1 T/T_n}~2 n\ \cos\frac{2\pi b}{n}\ ,
\label{jain-d}
\\
SU(2) {\rm\ NAF\ and\ } \ov{\mathrm{RR}}: && 
D_a^\ell = e^{-4\pi^2 h^1 T/T_n}~ (-1)^a\ 4 \cos\frac{\pi(\ell+1)}{k+2} \ ,
\label{arr-d}
\\
SU(n)\times {\rm Ising\ BS}: &&
D_{a,m} =  e^{-4\pi^2 h_1 T/T_{SU(n)}}~2 n\ \cos\frac{2\pi a}{n} + 
\nl
&& +\left\{
\begin{array}{ll}
(-1)^{a+m/2} \sqrt{2}\ e^{-\pi^2 T/(4T_{\rm Ising})}, & m=0,2,
\\
- e^{-2\pi^2 T/T_{\rm Ising}}, & m=1.
\end{array}
\right.
\label{s-shifts}
\ea
In all these expressions, we used the notations introduced earlier
in the respective sections two; $T_{SU(n)}$ and $T_{\rm Ising}$
are proportional to the neutral edge velocities of the corresponding terms
in the BS theory.
We remark that the difference between the Jain (Abelian) 
(\ref{jain-d}) and Read-Rezayi (non-Abelian) (\ref{rr-d}) correction
is precisely due to the neutral $S$-matrix term.

The corresponding $T>T_{n}$ Coulomb peak separations are,
\ba
&&SU(n) \ \mathrm{Jain}: 
\nl
& & \qquad\quad
\D\s_i =\frac{1}{\nu}   -\frac{T}{T_{ch}}
e^{-4\pi^2 h_1 T/T_n}\ 4n \cos \frac{2\pi i}{n}
\left(\cos \frac{2\pi}{n} - 1 \right) ,
\nl
&& {\rm\ NAF\ and\ } \ov{\mathrm{RR}}: 
\nl
& & \qquad\quad
\D\s^\ell_i =\frac{1}{\nu} -\frac{T}{T_{ch}} 
e^{-4\pi^2 h^1 T/T_n} (-1)^i\ 16 \cos\frac{\pi(\ell+1)}{k+2}  ,
\nl
 && SU(n)\times {\rm Ising\ \ BS}:
\nl
& & \qquad\quad
\D\s^m_i =\frac{1}{\nu} -\frac{T}{T_{ch}}
 e^{-4\pi^2 h_1 T/T_{SU(n)}}~4 n \cos\frac{2\pi i}{n} 
\left(\cos\frac{2\pi}{n} -1 \right)+ 
\nl
&& \qquad\qquad \qquad +\left\{
\begin{array}{ll}
(-1)^{i+m/2} 4 \sqrt{2}\ e^{-\pi^2 T/(4T_{\rm Ising})}, & m=0,2,
\\
0, & m=1.
\end{array}
\right.
\label{s-shift2}
\ea

Finally, the Coulomb peaks could be similarly predicted from the
partition function in the temperature range $T>T_{ch}$ by
performing the $S$ transformation on both $\t_n$ and $\t$.
However, the experimental values of $T_{ch}$ for small droplets
are comparable to the bulk gap $\D$ of Hall
states in the second Landau level, such that the CFT description 
is doubtful for higher temperatures.


\subsubsection{\bf $T>0$ off-equilibrium.}

In this section we discuss the $T>0 $ Coulomb blockade in presence of a
finite potential $\D V_o$ between the Hall droplet and
left (L) and right (R) reservoirs, leading to a steady flow of
electrons through the droplet.  This setting 
resembles a scattering experiment and is sensible to the
multiplicity of edge states $d^\ell_a$ discussed in the previous
section.

We shall study the problem within the phenomenological approach of the
Master Equation of Ref. \cite{noise}, that was successfully applied to
studying the fluctuations of the CB current in a quantum dot, in
particular its suppression due to Fermi statistics when the
transmission rates are not too small.  Clearly, a more precise
analysis of off-equilibrium physics would need the knowledge of the
finite-temperature real-time current-current correlation function that
is beyond the scopes of this work.

The starting point of \cite{noise} is the evolution equation 
of the semiclassical density matrix $\rho$, 
$d\rho/dt = M \rho$, where $M$ is the
matrix of transition rates. The latter can be written in the basis of states
with definite electron number $n$ inside the dot: its components
are the rates $\G_{ij}$ that can be computed by the Fermi golden rule 
and have the following factor for ideal-gas statistics: 
\be 
\g(\eps)=\frac{\eps}{1-e^{-\b \eps}}\ ,
\label{f-factor}
\ee
where $\eps=E_i-E_f$ is the energy of one-particle transitions.
This expression has two natural limits:
\be
\g (\eps) \sim \left\{
\begin{array}{lll}
\eps\ ,  && \b\eps \gg 1,
\\
\vert \eps \vert \ e^{-\b |\eps|}\ , && \eps < -T \ ,
\end{array} \right.
\label{eps-lim}
\ee
showing the $T\to 0$ phase-space enhancement for
allowed electron transition and the distribution of thermal-activated 
forbidden transitions, respectively.

The transition rates for one electron entering ($n\to n+1$) or leaving
($n\to n-1$) the dot, coming from the
(L) or (R) reservoirs are respectively given by \cite{noise}:
\be
\G^{\rm L(R)}_{n\to n\pm 1} = \frac{1}{e^2 R_{\rm L(R)}}\ 
\g\left[ \mp e \left(V-V_{\rm L(R)}\right) - E_C \right]\ ,
\label{g-expr}
\ee
where $R_ {\rm L(R)}$ and $V_{\rm L(R)}$ are the resistance and
potential levels of the reservoirs, and $V$ is the dot potential
(here we assume $V_L-V=V-V_R=\D V_o$). In the case of Hall
droplets, the Coulomb energy $E_C$ should be replaced by the difference of
edge energies 
$\D E_\s =E_\s(Q+1)-E_\s(Q)$ (\ref{def-en}), including neutral parts, 
that are obtained from the CFT partition function 
as seen before (we also set $e=1$ in the following).

Among the different regimes considered in \cite{noise}, that of $T<\D
V_o<\D E_\s$ is the most relevant for our purposes: this is the
situation of thermal-activated CB conduction, where few electrons can
enter the droplet from the left by thermal jumps and then quickly get
out to the right. The rate for the combined process is:
\be
\G=\G^L_{n\to n+ 1}\G^R_{n+1\to n}\sim \left(
(\D E_\s)^2 -(\D V_o)^2 \right) \ e^{-\b \left(\D E_\s-\D V_o \right)},
\label{g-tot}
\ee 
and the time interval between the peaks is $\D t \sim 1/\G$.

The main observation in this section is the following: 
if the Hall states possess multiplicities 
$d^\ell_a$ (\ref{na-mult}) for edge electron states,
the corresponding transition rates are amplified accordingly:
\be
\G \ \to \ d^\ell_a\ \G .
\label{ext-rate}
\ee

Therefore, a real-time experiment of peak rate can provide 
a direct test of edge multiplicities, and be useful to 
distinguish between Hall states with otherwise equal CB peak patters.
Clearly, the formula (\ref{g-tot}) depends on several unknown and
state-dependent phenomenological
parameters, such as reservoir-droplet couplings.
Nevertheless, the qualitative signal should be detectable:
upon parametric variation of $\s$ one can find the points 
of level matching $\s=\s_i$, 
where $\D E_\s\to 0$ and the rate $\G$ saturates.
From these points, one can tune $\s$ at the midpoints in-between
and test the formula (\ref{g-tot}) of thermal activated CB conduction
with less uncertainty. The signal is more significant
if the values of $d^\ell_a$ change considerably from one $\s$ interval to the
following (as e.g. for $SU(n)\times U(1)$ Jain states with 
$d_a=\left({n \atop a} \right)$).

\subsection{Thermopower and $T=0$ entropies}

Quasiparticles with non-Abelian statistics are characterized by
degenerate energy levels that are due to the multiplicity of fusion channels
\cite{stern-rev}. Their wave functions, described by 
Euclidean RCFT correlators in the plane, form multiplets whose dimension 
can be obtained by repeated use of the fusion rules,
$\phi_a\times\phi_b=\sum_c(N_a)_b^c\ \phi_c$, 
where $(N_a)$ is the fusion matrix.
The number of terms obtained by fusing $n_{qp}$ quasiparticles
of $a$-th type is given by:
\ba
{\cal D}_a(n_{qp})&=&\sum_{b=1}^N \left[\left(N_a\right)^{n_{qp}-1}\right]_a^b\ 
\nl
& \sim& d_a^{n_{qp}-1}\ , \qquad n_{qp}\ \to\ \infty,
\qquad\quad d_a=\frac{S_{a0}}{S_{00}}\ ,
\label{q-dim}
\ea where $d_a$ is the so-called quantum dimension (not to be confused with
other multiplicities discussed in the previous sections).  This result was
obtained by using the Verlinde formula \cite{cft} for diagonalizing the fusion
matrix and by keeping the contribution of the largest eigenvalue for
$n_{qp}\to\infty$.

From the thermodynamical point of view, the presence of $n_{qp} \gg 1$
quasiparticles in the system implies a quantum entropy at $T=0$, 
\be 
{\cal S}_a(T=0)\sim n_{qp}\log\left(d_a \right)\ , 
\label{ent-0}
\ee 
that is ``unexpected'' because the state is completely determined
(gapped, fixed control parameters, fixed quasiparticle positions,
etc.).

This entropy can be obtained from the RCFT partition function described in
this paper, that encodes all the low-energy physics of static bulk
quasiparticles and massless edge excitations.  Let us consider an isolated
droplet of Hall fluid, i.e. the disk geometry, and compute the entropy first
for one quasiparticle and then for many of them.

The disk partition function for a quasiparticle
of type $a$ in the bulk is given by the generalized character 
$\theta_a(\tau)$ (cf. (\ref{z-disk}) and section 2.1), owing to the 
condition of global integer charge,
$Q_{\rm bulk}+Q_{\rm edge}\in\Z$.
The $T=0$ entropy is a many-body effect that manifests itself in
the thermodynamic limit; therefore, we should first send $R\to\infty$
 and then $T\to 0$, i.e. expand the partition function for $\t\to 0$
\cite{cft}.
Upon performing the $S$ modular transformation (of both neutral and
charged parts), we obtain the leading behavior, 
$\theta_a(\t)\sim S_{a0}\ \theta_0(-1/\t)\sim S_{a0} \exp(i2\pi c/(24 \t))$, 
and compute the entropy:
\be
{\cal S}_a(T\to 0)=\left(1-\t \frac{d}{d\t} \right)
\log\theta_a \sim \log\frac{S_{a0}}{S_{00}} -\log \frac{1}{S_{00}},
\label{therm-e}
\ee 
The first contribution indeed reproduces the entropy for one 
non-Abelian quasiparticle, ${\cal S}_a=\log d_a$;
note that the Abelian part of the $S$-matrix cancels out
(e.g. $S^{\rm (Abelian)}_{0a}=1/\sqrt{\hat p}$, $\forall a$, 
in the Read-Rezayi state)
and one obtains the quantum dimension (\ref{q-dim}).

The last term in (\ref{therm-e}) is the (negative) boundary
contribution to the entropy, that is also present without
quasiparticles; it involves the ``total quantum dimension'',
${\cal D}=1/S_{00}=\sqrt{\hat p}\sqrt{\sum_a d^2_a}$ 
\cite{tee2}, where the sum extends over all quasiparticles, and
receives a contribution from the Abelian part.  Note that a single
quasiparticle is not usually associated to a bulk entropy, as
described at the beginning of this section, because multiple fusion
channels only appear for $n_{pq}\ge 4$ quasiparticles.  On the other
hand, from the topological point of view, the edge divides the infinite
system in two parts, interior and exterior, with the edge keeping
track of the missing part \cite{tee2}.

The entropy for several quasiparticles in the bulk can be obtained
from the corresponding disk partition function; for two
quasiparticles, for example, this is obtained by fusing the two
particles and summing over the resulting edge sectors, as follows: 
\be
Z_{aa} = \sum_b N_{aa}^b\ \theta_b(\t)\ .
\label{s-two}
\ee
In general, the repeated fusion of several particles reproduces 
the computation of the bulk entropy contribution 
(\ref{q-dim}), leading again to ${\cal S}\sim n_{qp}\log (d_a)$.

Several aspects of the non-Abelian entropy have been discussed in
\cite{tee}\cite{tee2}; here we deal with the proposal of observing it
in thermopower measurements \cite{thermop}, that could be feasible
\cite{thermo-exp}.  Let us consider the annulus
geometry and introduce both an electric potential difference $\nabla
V_o$ and a temperature gradient $\nabla T$ between the two edges. 
The electric current takes the form: 
\be 
{\bf J}= -\sigma \cdot \nabla V_o - \a \cdot \nabla T\ ,
\label{j-def}
\ee
where $\sigma,\alpha$ are the electric and Peltier conductivity tensors.
The thermopower (or Seebeck coefficient) is defined as the tensor, 
$S_{\rm Seebeck} =\sigma^{-1}\cdot \a$, i.e. the ratio of transport
coefficients pertaining to the two gradients \cite{cooper}.

Here we consider the case of exact compensation between the gradients,
such that the current vanishes, ${\bf J}=0$. In this case, 
the thermopower component in the annulus geometry is given by: 
\be S_{\rm Seebeck} = \frac{\a}{\s} =
-\ \frac{\D V_o}{\D T}\ = \frac{\cal S}{eN_e}\ .
\label{therm-p}
\ee
In the last part of this equation, we also wrote the desired relation,
that the thermopower is equal to the entropy per electron  
\cite{cooper}\cite{thermop}.

In the following, this result is recovered by adapting 
the near-equilibrium description by Yang and Halperin \cite{thermop}
to our setting.
We consider a small variation of the grand-canonical potential, 
\be d\Omega = -{\cal S} dT -N_e(d\mu+e~ dV_o) ,
\label{g-can}
\ee 
involving both the chemical $\mu$ and
electric $V_o$ potentials coupled to the $N_e$ electrons.  Note that
Eq. (\ref{g-can}) takes the non-relativistic form because we are
describing bulk effects related to adding the quasiparticles.
For vanishing current, the gradients induce an excess of
charge at the edges that is equivalent to the pressure effect considered
in \cite{thermop}.
Therefore, from the definition (\ref{j-def}) and the grand-canonical potential 
(\ref{g-can}), we can express the two conductivities in terms of
second derivatives (implying the Maxwell relations), as follows:
\be
\s=-\frac{\partial Q}{\partial V_o}=
e^2\frac{\partial^2\Omega}{\partial\mu^2}=
-e^2\frac{\partial N}{\partial \mu}\ ,
\qquad
\a=-\frac{\partial Q}{\partial T}=
e\frac{\partial^2\Omega}{\partial\mu \partial T}=
-e\frac{\partial \cal S}{\partial \mu}\ .
\label{II-der}
\ee
Upon taking their ratio, the result (\ref{therm-p}) is recovered
for ${\cal S}$ linear in the number of electrons, as shown momentarily.

From the experimental point of view, quasiparticles of smaller charge
($a=1$) are induced in the system by varying infinitesimally the magnetic
field from the center of the plateau $B=B_o$.
In the diluted limit, $\nu$ remains constant and the number of 
quasiparticles $n_{qp}$ and of electrons can be related as follows:
\be
n_{qp}= \frac{e(B-B_o)}{e^* B_o}\ N_e\ . 
\label{n-expr}
\ee 
The entropy associated to the non-Abelian quasiparticles is 
given by partition function as explained before: in the 
annulus geometry, its expression for $T\to 0$ is given by the one-edge
expression, e.g. (\ref{s-two}), multiplied by the ground-state
contribution for the other edge $\ov{\theta(\tau)}_0$.
The result is again given by (\ref{ent-0}) up to a constant.

Finally, the value of the thermopower is found by taking the ratio 
(\ref{therm-p}) of the entropy over the total electron charge \cite{thermop}: 
\be 
S_{\rm Seebeck} =
\left\vert\frac{(B-B_o)}{e^* B_o}\right\vert \log(d_1)\ .
\label{therm-r}
\ee 
Upon measuring the two gradients $\nabla T,\nabla V_o$ near the
center of the plateau, one can observe a characteristic V-shaped
behavior of $S_{\rm Seebeck}$ signaling the non-Abelian state
\cite{thermo-exp}; other sources of entropy are under control in the
gapped state.

In conclusion, we have shown that the RCFT partition function of edge
excitations is useful to obtain the $T=0$ entropies associated to
non-Abelian quasiparticles.

\section{Conclusions}

In this paper, we have obtained the modular invariant partition
functions of several non-Abelian states that could describe the Hall plateaux
in the second Landau level. We have extended and
simplified the earlier derivations \cite{cz}\cite{cgt2}\cite{cvz}
and showed that they straightforwardly follow from the choice of the RCFT 
for neutral states and of the electron field, corresponding to
a so-called simple current.

The physical applications to Coulomb blockade experiments \cite{cb-exp}
have been discussed in section 4: for $T>0$ there are two corrections to
the periodic peak positions found at $T=0$ \cite{cb0}, 
in the ranges $T<T_n$ and $T_n<T<T_{ch}$,
respectively, where $T_n$ and $T_{ch}$ 
are typical energies of neutral and charged excitations ($T_{ch}\sim 10\ T_n$).
These corrections are sensible to the multiplicity of low-lying neutral states 
\cite{georgiev}
and to their $S$-matrix of modular transformation \cite{cbT}: therefore,
they can give a richer and unambiguous signal of non-Abelian statistics
\cite{dopp}.

Using a phenomenological approach, in section 4.1.3 we also argued that
the multiplicity of neutral states could be better seen in experiments
observing the Coulomb-peak time rate off-equilibrium, 
i.e. at finite bias $V_0>0$.

Finally, partition functions were used to compute the
thermopower and the associated entropy of non-Abelian quasiparticles;
here, earlier topological approaches were recovered and extended by
using the physical partition function of the Hall system with an edge
in the $T=0$ limit.

A new perspective in our RCFT approach based on annulus partition
functions is offered by the possibility of finding other solutions
of the modular invariant conditions, where the neutral excitations
of the left and right edges are paired differently.
Actually, the $U$ condition (\ref{u-cond}) requires the matching
of fractional charges on the two edges, but leaves the possibility 
on non-trivial pairing of neutral states, still constrained by the
other modular conditions. 

For example, if the neutral CFT possesses a second simple current, the
chiral symmetry can be further extended by orbifolding; the resulting
partition function is diagonal with respect to the new extended-symmetry basis
but is non-diagonal in the original basis.
The new theory possesses less excitations than the original one
and an additional neutral excitation (the second simple current).
Such a solution is possible for example in the Read-Rezayi states with
$k\ge 4$ even: its physical interpretation remains to be understood.

Non-diagonal modular invariants has been much analyzed in
the RCFT literature \cite{cft}: in the application to critical phenomena in
two dimensions, they describe new universality classes that are
different from those of diagonal invariants \cite{ade}.  It would be
interesting to study this kind of model building in the quantum Hall setting
because it may reveal new non-Abelian features.

\ack

We thanks D. Ferraro,  N. Magnoli, I. D. Rodriguez, M. Sassetti,
A. Stern and G. R. Zemba for interesting discussions.  
A.C. would like to thank the hospitality of LPTHE, ENS, Paris. This work
was partially supported by the ESF programme INSTANS and by a PRIN grant of
Italian Ministry of Education and Research.

\appendix

\section{Modular transformations}

In this Appendix, we collect some definitions, properties and
modular transformations of the non-Abelian characters used in the text.
A brief review on modular invariance and modular forms can be found
in \cite{cvz} and the reference therein; more extensive material is 
presented in \cite{cft}.

\subsection{Read-Rezayi}
\label{rr-app}

The properties and transformations of parafermion characters
(\ref{pf-char}) and  (\ref{pf-s}) are obtained 
from the coset construction \cite{cft}\cite{qiu}\cite{gepner}.
The basic identity of the coset 
$\wh{SU(2)}_k/\wh{U(1)}$ (resp. $\wh{SU(3)}_k/\wh{U(1)}^2$ for the NASS state)
is the following expansion of the affine $\wh{SU(2)}_k$ 
(resp. $\wh{SU(3)}_k$) characters $\c^\L$:
\begin{equation}
\label{defchi}
\chi^\L(\t,\z)=\sum_{\l\in P/kQ}\chi^\L_\l(\t)\ \vartheta_\l(\t,\z)\,,
\end{equation}
where $\chi^\L_\l$ are the parafermionic characters and $\vartheta_\l$
are classical theta function at level $k$
associated to the root lattice $Q$ of the Lie
algebra \cite{cft}. The indices $\L$ and $\l$ belong to the weight lattice
$P$ (cf. Section 3.4).

For the $SU(2)$ lattice,
$\vartheta_{m/\sqrt{2}}(\t,\z)=K_{m}(\t,k\z;2k)$, the Abelian theta
function (\ref{thetaf}), and $\chi^\L_\l=\chi^\ell_m$ the $\Z_k$
parafermion characters.  From (\ref{defchi}) we also obtain the
embedding index for the cosets, $2k$ and $2k\times 6k$ for
$\wh{SU(2)}_k/\wh{U(1)}$ and $\wh{SU(3)}_k/\wh{U(1)}^2$, respectively.
The modular transformations of the characters $\chi^\ell_m$ are
determined in such way that (\ref{defchi}) reproduces the correct
transformation of the $\wh{SU(2)}_k$ characters:
\be
 \chi^\ell\left(\frac{-1}{\t}\right)
=\sum_{\ell'=0}^{k}\
s_{\ell,\ell'}\ \c^{\ell'}(\t)\ ,\quad
s_{\ell,\ell'} = \sqrt{\frac{2}{k+2}} \ 
\sin\frac{\pi (\ell+1)(\ell'+1)}{k+2} .
\label{su2k}
\ee
(For simplicity, we fix $\z=0$ in neutral characters).
The combination of (\ref{defchi}) and (\ref{su2k})
yields the transformation of the characters $\chi^\ell_m$ in (\ref{pf-s}). 

The field identifications (\ref{pf-char}),(\ref{f-id}), leading to
symmetries among the parafermionic characters, follow  
from the properties of the modular transformations \cite{gepner}. 
For example, in the $\wh{SU(2)}_k$ case the matrix $s_{\ell,\ell'}$ 
obeys the following symmetry under $\ell\to  {\cal A}(\ell)=k-\ell$:
\be 
\label{su2A}
{\cal A}\left(s_{\ell^\prime,\ell}\right) \equiv
s_{{\cal A}(\ell^\prime),\ell} =s_{k-\ell^\prime,\ell} =
(-1)^\ell\ s_{\ell^\prime,\ell}\,.  
\ee

We now describe the modular transformation (\ref{pf-th-s}) 
of the Read-Rezayi extended characters $\theta^\ell_a$ (\ref{pf-th}).
After transformation of each term in their sum,  
the sum over the running index $b$ yields:
\be
\theta^\ell_a(-1/\t) = \frac{k}{\sqrt{2~k~p}}\sum_{ q'=0}^{p-1}
\sum_{m^\prime=0}^{2k-1}\sum_{\ell=0}^{k}\delta^{(k)}_{m',q'}\ 
e^{2 \pi i \frac{2 a  q'- \hat{p}a m' }{ p }} 
s_{\ell,\ell^\prime}K_{q'}\c^{\ell'}_{m^\prime}\,.
\label{pf-th-sA}
\ee

The mod-$k$ delta function is solved by $m'=q'+\sigma k$ with $\s=0,~1$;
then, $q'$ is re-expressed as $q'=a'+b'\hat{p}$,
with $a^\prime$ mod $\hat{p}$ and $ b'$ mod $k$. 
The sum on $q'$ and $m^\prime$ can be decomposed into
$\sum_{\s=0}^1\sum_{a'=0}^{\hat{p}-1}\sum_{b'=0}^{k-1}$, and the phase is 
rewritten,  $(-1)^{\s a}\ e^{-2 \pi i \frac{M a a'}{2\hat{p}}}$. 
Using the identification (\ref{pf-char}) and (\ref{su2A}), the sum on $\s$
becomes $2\delta^{(2)}_{a,\ell}$ and finally the result 
(\ref{pf-th-s}) is obtained. 

Calling the whole $S$-transformation of the $\theta^\ell_a$ characters
in (\ref{pf-th-s}) as 
$S^{\ell~\ell^\prime}_{a,a'}$, we now check its unitarity.
We find that,
\be
\label{pf-SS1}
  \sum_{\ell',a'}S^{\ell\ell'}_{a,a'}( S ^\dagger)^{\ell'\ell''}_{a',a''}
=\frac{1}{\hat{p}}\sum_{\ell'=0}^k\delta_{a,\ell}^{(2)}\delta_{a'',\ell''}^{(2)}
s_{\ell~\ell'}s_{\ell'~\ell''}
\sum_{a'=0}^{\hat{p}-1}e^{2\pi i \frac{a'(a-a^{\prime\prime})M}{2\hat{p}}} \ .
\ee
Since the non-Abelian part is unitary, we obtain 
$\delta^{(2)}_{\ell,\ell''}$ and thus $\d^{(\hat p)}_{(a-a'')M/2,0}$;
the latter condition is equivalent to $\d^{(2\hat p)}_{aa''}$ because
$\hat p$ and $M$ are coprime, $(\hat p,M)=1$.

\subsection{NAF and $\ov{\mathrm{RR}}$}

We now derive the modular transformation  of the
NAF characters $\theta^\ell_a$ in (\ref{su2-theta}). The transformation of
the two terms in their expression leads to:

\begin{equation}\label{su2-sA}
  \theta^\ell_a(-1/\t)=\frac{1}{\sqrt{2\hat{p}}}
\sum_{q'=0}^{2\hat{p}-1}\sum_{\tilde{\ell}=0}^k e^{2 \pi i \frac{aq'}{2\hat{p}}}
\left(s_{\ell~\tilde{\ell}}+s_{k-\ell~\tilde{\ell}}e^{ \pi i q'}\right)
\c^{\tilde{\ell}} K_{q'}\,.
\end{equation}
The term in the parenthesis is 
$2s_{\ell~\tilde{\ell}}\delta^{(2)}_{q',\tilde{\ell}}$, using
(\ref{su2A}); this condition can be solved by the following
parameterization: $q'=a'+b'\hat{p}$ and
$\tilde{\ell}=\ell'$ for $b^\prime=0$, and 
$\tilde{\ell}=k-\ell'$ for $b^\prime=1$, with
$a'=0$ mod $\hat{p}$ and $\ell^{\prime}=a^{\prime}$ mod $2$. The
sums on $q ^{\prime}~\mathrm{and}~ \tilde{\ell}$ become sums on
$a^{\prime},~\ell^{\prime}$ and $b^{\prime}$. The sum on $b^{\prime}$ is,
$$ 
e^{2\pi i\frac{aa^\prime}{2\hat{p}}}
\left(s_{\ell~\ell^{\prime}}K_{a^\prime}\chi^{\ell^\prime}+
s_{\ell~k-\ell^{\prime}}e^{i\pi a}\chi^{k-\ell^\prime}K_{a'+\hat{p}}
\right)=
$$
$$
=e^{2\pi i\frac{aa^\prime}{2\hat{p}}}  
\delta^{(2)}_{a~\ell}s_{\ell~\ell^{\prime}} 
\left(K_{a^\prime}\chi^{\ell^\prime}+\chi^{k-\ell^\prime}K_{a'+\hat{p}}
\right)\, ,
$$
finally leading to (\ref{su2-s}).

The unitary of the modular transformation in (\ref{su2-s}) can be
verified following the same steps of the previous section.
For $\ov{\rm RR}$ fluids the computation is the same due to the reality of
the $\wh{SU(2)}_k$ $S$-matrix; a little difference is that $k+M$ is odd
for NAF and $M$ is odd for $\ov{\mathrm{RR}}$.

\subsection{ Bonderson-Slingerland states}

The BS case is easier by changing the basis of characters 
from (\ref{bs-char2} to the one in (\ref{bs-s1}), because the new
Ising characters possess simpler transformations. This new basis is
($a=0,1,\dots,nM$):
\ba
\wt{\theta}_{a,0}=&
\frac{1}{\sqrt{2}}\left(\theta_{a,0}+\theta_{a,2}\right)
=\sum_{b=1}^{2n} K_{2an+b\hat{p}}\left(\t,2n\z;2n\hat{p} \right)
\ \Theta_b ~\wt{\chi}_0  , \ m=0 
\nl
\wt{\theta}_{a,1}=&
\theta_{a,1}=\sum_{b=1}^{2n} 
K_{(2a+1)n+b\hat{p}}\left(\t,2n\z;2n\hat{p} \right)
\ \Theta_b\ \wt{\chi}_1 , \qquad\qquad m=1,
\nl
\wt{\theta}_{a,2}=&
\frac{1}{\sqrt{2}}\left(\theta_{a,0}-\theta_{a,2}\right)
=\sum_{b=1}^{2n} e^{i\pi b}K_{2an+b\hat{p}}\left(\t,2n\z;2n\hat{p} \right)
 \Theta_b \wt{\chi}_2 , \  m=2,
\label{bs-char3}
\ea
The computation requires the modular transformation of $\widehat{SU(n)}_1$
characters \cite{cz}:
\be
\label{s-sun1}
 \Theta_b \left(- \frac{1}{\tau}\right) = 
 \frac{1}{\sqrt{n}}\sum_{b'=1}^n\ {\rm e}^{-i2\pi\frac{bb^\prime}{n} } 
\ \Theta_{b^\prime}\left(\tau\right)\ . 
\ee
Combining all the transformations in the factors of $\wt{\theta}_{a,i}$,
we write:
\ba
\label{bs-sA}
\wt{\theta}_{a,i}&=&\frac{1}{\sqrt{n^2 2 \hat{p}}}\sum_{b=1}^{2n}
\sum_{q^\prime=1}^{p}\sum_{\b^\prime}^{n}
e^{2\pi i \frac{(2an+n\delta_{i,1}+b\hat{p})q^\prime-2\hat{p}b\beta^\prime}
{2n\hat{p}}+\pi i \delta_{i,2}b} \times
\nl
&&\times K_{q^\prime}\Theta_{\b^\prime}\ 
\sum_{i'=0}^2S^{\mathrm{Ising}}_{i,i^\prime}\ \wt{\c}_{i^\prime}\,,
\ea
where the modular transformation,
$$S^{\mathrm{Ising}}=\left(\begin{array}{ccc}
1&0&0\\0&0&1\\0&1&0
\end{array}
\right),
$$
is defined according to (\ref{bs-s1}).
The sum on $b$ gives the factor $\delta^{(2n)}_{q', 2\b'+n\delta_{i,2}}$, whose
solution requires to parameterize $q^\prime$ as 
$q^\prime=r^\prime+\wt{b}\hat{p}$, with
$r^\prime$ mod $\hat{p}$ and $\wt{b}$ mod $2n$. The 
delta function imposes,
$$
r^\prime-n\delta_{i,2}+2\wt{b}-2\b^\prime=2 l n\,, \qquad
~\mathrm {i.e.} \qquad
r^\prime-n\delta_{i,2}=2a^\prime\,,\qquad
\wt{b}-\b^\prime=\sigma n\,,
$$
since $\wt{b}$ and $\b^\prime$ are defined mod $2n$ and $n$,
respectively, $\s=0,~1$. It is convenient to write
$r^\prime=2a^\prime+n\delta_{i,2}$, with $a^\prime$ 
mod $\hat{p}/2$, which is coprime with $n$; therefore, we can
write $r^\prime=2a^\prime n+n\delta_{i,2}$. 
Note that $\b^\prime$ is the index mod $n$ of $\Theta_{\b^\prime}$, 
thus we can replace it with $\wt{b}$. After these substitutions 
in (\ref{bs-sA}), we obtain the modular
transformation of the Bonderson-Slingerland characters
(\ref{bs-char3}), with $S$  reported in (\ref{bs-s2}).  
The unitarity of this matrix is:
\ba
&&\sum_{a',m'}S_{(a,m),(a',m')}S_{(a',m'),(a'',m'')}^\dagger=\sum_{a'}
e^{2 \pi i\frac{a^\prime(a-a^{\prime\prime})}{mM+1}}\times
\nl
&&\left(
\begin{array}{lll}
1 & 0 & 0 \\
0 & 0 & e^{i2\pi a'n/2(nM+1)} \\
0 & e^{i2\pi an/2(nM+1)} & 0
\end{array}
\right)\times
\nl
&&\left(
\begin{array}{lll}
1 & 0 & 0 \\
0 & 0 &e^{-i2\pi an/2(nM+1)}  \\
0 & e^{-i2\pi a'n/2(nM+1)} & 0
\end{array}
\right)=\delta^{(mM+1)}_{a~a^{\prime\prime}}\delta_{m~m^{\prime\prime}}\,.
\ea

\subsection{NASS states}

The modular transformations of the $\wh{SU(3)}_k/\wh{U(1)}^2$
parafermion characters is found from (\ref{defchi}): the
$S$-matrix of the $\wh{SU(3)}_k$ characters in the numerator of the
coset is,
\ba
&&\chi^\L \left(\frac{-1}{\t}\right)=\sum_{\L^\prime\in
  P^+_k}s_{\L\L^\prime}\ \chi^{\L'}(\t) , 
\nl
&&s_{\L\L^\prime}=\frac{i}{\sqrt{3}(k+3)}\sum_{w\in W} (-1)^{|w|} 
\exp\left[2\pi i 
\frac{ \left( w( \Lambda+\rho)~,~\Lambda^\prime+\rho \right)}{k+3}
\right],
\ea
where $w$ is an element of the Weyl group $W$ of $SU(3)$, $|w|$ is its length, 
and $\r$ is half of the sum of the positive roots \cite{cft}.
The coset decomposition yields the following transformations 
of the parafermionic characters:
\be
\c^{\L} _{\l} (-\frac{1}{\t})  =\frac{1}{\sqrt{3}k}
\sum_{ \l^\prime\in P/kQ} e^{- 2 \pi i\frac{(\l, \l')}{k}} 
\sum_{ \L^\prime \in P^k_+} s_{\L\L^\prime}\ \c^{ \L^\prime} _{\l^\prime} (\t)
\label{pfk3}\, ,
\ee
where the ranges of the indices are explained in section 3.4.
Useful properties of $s_{\L\L'}$ are its transformations under 
the automorphism ${\cal A}$ \cite{gepner}:   
\ba
\label{su3A}
&&\L=(n_1,n_2)\mapsto {\cal A}(\L)=(k-n_1-n_2,n_1)\ , 
\nl
&& \qquad
{\cal A}\left(s_{\L^\prime,\L}\right) \equiv s_{{\cal A}\L^\prime,\L} 
=s_{\L^\prime,\L} ~e^{2 \p i \frac{2 n_1+n_2}{3}};
\ea
this is the $\wh{SU(3)}_k$ version of (\ref{su2A}): note that
this map obeys ${\cal A}^3=1$. 
The field identifications (\ref{f-id}) are
deduced by studying the the modular matrix of the
parafermionic characters \cite{gepner}. 
Equation (\ref{f-id}) can be rewritten as 
$\chi^\L_\l=\chi^{{\cal A}(\L)}_{{\cal A}(0)+\l}=
\chi^{{\cal A}^2(\L)}_{2{\cal A}(0)+\l}$.

In order to derive the modular transformation of NASS characters
(\ref{nass-s}), we rewrite (\ref{nass-th2}) as follows:
\be
\label{th2-nass}
\Theta _{q,s}^{n _1 n _2}\left(\tau ,\zeta \right)=
\sum _{a,b=0}^{k-1} \sum _{\iota=0,1}K_{q+\hat{p}(a+b)+\iota k\hat{p}}^{(Q)}~
K^{(S)}_{s+(a-b)+\iota k}~
\chi _{\frac{q+s}{2}+2a+b,\frac{-q+s}{2}-a-2b}^{n _1,~n _2}\,.
\ee
After transformation of each term in this sum, we find 
\ba
\label{S12NASS}
&&\Theta _{q,s}^{n _1n _2}\left(-\frac{1}{\tau}\right)=
\frac{1}{\sqrt{2kp}}    \sum _{\iota=0,1}\sum_{a,b=0}^{k-1}
\sum _{s^\prime=0}^{2k-1}\sum _{c^\prime=0}^{p-1}\sum_{\mu\in\frac{P}{k Q }} 
\sum_{ {\L^\prime} \in P^k_+} \ S_{ \L \L'}  
\nl
&&\times\exp\left[2\p i \left(
\frac{(m+\hat{p}(a+b)+\iota k\hat{p})c^\prime}{q}+
\frac{(s+(a-b)+\iota k)s^\prime}{2k}-
\frac{(\l, \m)}{k}\right)\right]
\nl
&&\times K_{c^\prime}^{(Q)}\ K^{(S)}_{s^\prime}
\ \chi _{\mu}^{ \L^\prime}(\tau ) 
\ea
where $\l$ is an abbreviation for the subscript of
$\c^\L_{\l}$ in (\ref{th2-nass}). The sum on $\iota$ gives
$2\delta^{(2)}_{c',s'}$. The next step requires to explicit the
product $(\l,\mu)$; then, the sums on $a$ and $b$ give the
conditions:
\be
\label{nass-delta}
\frac{ c^\prime+ s^\prime}{2}-\m_1=0\quad\mathrm{mod} ~k\,, \qquad\qquad
\frac{ c^\prime- s^\prime}{2}+\m_2=0\quad\mathrm{mod} ~k\, .
\ee
These equations have three solutions in the fundamental domain
$\mu\in P/kQ$ that should be taken into account.
As before, we reparameterize the
indices $c^\prime$ and $s^\prime$:
\be
c^\prime=\tilde{q}+\hat{p}(\tilde{a}+\tilde{b})+\tilde{\iota}k\hat{p}\,, 
\qquad s^\prime=\tilde{s}+(\tilde{a}-\tilde{b})+\tilde{\iota}k\, . 
\ee
Note that since $c^\prime =s^\prime$ mod 2 also $\tilde{q}=\tilde{s}$ mod 2. 
In this parameterization, the three solutions of (\ref{nass-delta}) are, 
\be
\mu_1=\frac{\tilde{q}+ \tilde{s}}{2}+2kN(\tilde{a}+\tilde{b})+
2 \tilde{a}+\tilde{b}+\tilde{n}k \ ,\qquad
\mu_2=\frac{-\tilde{q}+\tilde{s}}{2}-2 \tilde{b}-\tilde{a}-\tilde{n}'k\ ,
\label{mu1}
\ee 
where the possible values of $(\tilde{n},\tilde{n}')$ are $(0,0)$,
$(1,0)$ and $(1,1)$. The weight $\mu$ in (\ref{mu1}) can be rewritten,
 \be
 \mu=\tilde{\mu}+tA(0)\ ,\qquad 
t=\tilde{n}+\tilde{n}'+2N(\tilde{a}+\tilde{b})\,.
\ee
Using the automorphism  (\ref{su3A}), we find,
$$
S_{\L \L^\prime}\
\chi_{\tilde{\mu}+tA(0)}^{\L'}=S_{ \L \L^\prime}\
\chi_{\tilde{\mu}}^{A^{-t}(\L')}= S_{\L A^{t}(\L'')} \ 
\chi_{\tilde{\mu}}^{(\L'')}=e^{2\pi i  t(2n_1+n_2)/3}S_{\L\L^{\prime\prime}}\
\chi_{\tilde{\mu}}^{\L''}\ ,
$$
with $\L''=A^{-t}(\L')$. 
Upon substituting in (\ref{S12NASS}), the phase becomes,
$$
2\pi i\left(-\frac{q \tilde {q} N}{ 3 p} - 
\frac{q}{6}(\tilde{n}+\tilde{n}')-
\frac{s}{2}(\tilde{n}-\tilde{n})+
\frac{qN(\tilde{a}+\tilde{b})}{3}+
\tilde{\iota}\frac{q+s}{2}+\frac{t(2 n_1+n_2)}{3}\right).
$$
The sum on $\tilde{\iota}$ yields $\delta^{(2)}_{q,s}$. 
The sum on the three values of $(n,n^\prime)$ leads to:
\be
\delta^{(3)}_{-\frac{q+3s}{2}+2 n_1+n_2,0}=
\delta^{(3)}_{n_1-n_2,q},
\label{deltamod3}
\ee
which is  the triality condition. Finally, 
the result in (\ref{nass-s}) is recovered.

The unitary of the the $S$-matrix is:
\ba
\label{nass-ss}
&&\sum_{q^{\prime}\L^\prime }S_{qq^\prime,ss^\prime}^{\L\L^\prime}
(S^\dagger)_{q' q'',s^{\prime}s^{\prime\prime}}^{\L^{\prime} \L^{\prime\prime}}=
\nl
&&=
\delta^{(3)}_{n_1-n_2,q}
\delta^{(3)}_{n_1^{\prime\prime}-n_2^{\prime\prime},q^{\prime\prime}}
\delta^{(2)}_{q,s}\delta^{(2)}_{q^{\prime\prime},s^{\prime\prime}}
\sum_{\L^\prime}s_{\L\L^\prime}\ s_{\L'\L''}^\dagger
\sum_{q^\prime}
e^{-2\pi i M\frac{q^{\prime}(q-q^{\prime\prime})}{3\hat{p}}}\,.
\ea
This expression is zero for $n_1-n_2\neq n_1''-n_2''$ mod $3$.
Then, for $n_1-n_2= n_1^{\prime\prime}-n_2^{\prime\prime}$, 
i.e. $\L-\L^{\prime\prime}\in Q$, the two delta mod $3$ impose 
$q-q^{\prime\prime}=3 n$ for an integer $n$;
the sum on the two indices $q^{\prime}$ and $\L^{\prime}$ plus the
unitary of $s_{\L~\L^\prime}$ show that the l.h.s. of (\ref{nass-ss}) is equal
to
$\delta_{(q-q^{\prime\prime})N/3,0}^{(\hat{p})}\delta_{\L,\L^{\prime\prime}}$.
For $N\neq 0$ mod $3$,
$\delta_{(q-q^{\prime\prime})N/3,0}^{(\hat{p})}=
\delta_{q-q^{\prime\prime},0}^{(\hat{p})}$,
since $(\hat{p},N)=1$, thus proving unitarity. Note that the case $N=1$
includes the physically relevant fractions, $\nu=2+\frac{4}{7}$ and
$\nu=2+\frac{2}{3}$, for $k=2,3$.

 
\end{document}